\documentclass{article}

\usepackage[letterpaper,top=2cm,bottom=2cm,left=3cm,right=3cm,marginparwidth=1.75cm]{geometry}
\usepackage{booktabs}

\usepackage{amsmath}
\usepackage{graphicx}
\usepackage{amsfonts}
\usepackage[colorlinks=true, allcolors=blue]{hyperref}
\usepackage[backend=biber,style=numeric,citestyle=nature]{biblatex}
\title{Neural Recording Power Optimization Through Machine Learning Guided Resolution Reconfiguration}

\author{Aviral Pandey\textsuperscript{*,1}, Dhruv Vaish\textsuperscript{*,1}, I-Ting Lin\textsuperscript{*,1}, Hanna Schlegel\textsuperscript{1}, Preeya Khanna\textsuperscript{1},\\ Rikky Muller\textsuperscript{1,2}\\
\textsuperscript{1}University of California, Berkeley, Berkeley, CA, USA,\\ \textsuperscript{2}Weill Neurohub, Berkeley, CA, USA \\
\textsuperscript{*} Equal Credit Authors}

\begin{document}
\maketitle

\begin{abstract}
\textcolor{black}{As neural interfaces scale to thousands of channels, they enable increasingly sophisticated decoding for both research and clinical applications. But channel counts must grow without raising total power, and because the amplification and digitization circuits are approaching theoretical efficiency limits, further reduction demands system-level innovation.  This work demonstrates resolution reconfiguration, in which the resolution of less important channels, as extracted from the machine-learning based decoders, is scaled down to save power. The methodology is simulated offline across three datasets: seizure detection from iEEG, gesture recognition from sEMG, and force regression from ECoG. With linear decoders, resolution reconfiguration consumes between 12x to 23x lower power than a traditional array. It also consumes between 3x to 5x lower power than channel selection, a prior approach in which less important channels were turned off. Greater savings were found for gesture recognition and force regression due to higher spatial distribution of information across electrodes. Resolution reconfiguration is further validated in a five-channel online cursor-control study from sEMG, where it saves 3.1x over a traditional array and 2.4x more than channel selection. The results demonstrate the power benefits of resolution reconfigurable front-ends and their wide applicability to neural decoding problems.}
\end{abstract}


\section*{Introduction}

Neural recording systems measure electrical signals generated in different regions of the nervous system. These signals are fundamental to how neurons communicate, process information, and regulate bodily functions. While optical and magnetic readout modalities have been demonstrated \cite{ebner1995use,caruso2017vivo}, many neural recording devices record electrical signals, such as local field potentials (LFPs) \cite{andy2018wand, steinmetz2021neuropixels, blackrockUtah, sahasrabuddhe2021argo}, electroencephalography (EEG) \cite{buzsaki2012origin, sabio2024eeg}, or electromyography (EMG) \cite{andy_emg_nature}. These devices are used for both neuroscience research \cite{paulk2022large} and clinical applications \cite{lee2015single} such as motor or communication prostheses \cite{daly_braincomputer_2008}, seizure monitoring implants, and closed-loop deep brain stimulators. 

Figure \ref{fig:figure1} illustrates a representative neural‐recording system. Application-specific electrodes, positioned either intracranially or extracranially, capture neural signals as voltages that are then amplified and digitized by an analog front-end (AFE). After digitization, the data may be processed on-device or streamed to an external computer for decoding. Clinically relevant metrics such as motor intent or seizure onset are inferred from these neural signals. Although the specific decoding pipeline depends on the target application, it generally comprises signal processing and feature extraction, followed by classification or regression via machine‐learning techniques (e.g. logistic regression, random forests, or neural networks) \cite{subasi2005classification, ayinala2012low, glaser2020machine}.

As decoding tasks grow in complexity, neural recording systems must integrate together more front-ends to support higher channel counts while also operating within tight power constraints imposed by battery life and tissue heating limits \cite{wolf_thermal_2008}. With data from more channels, larger models will enable new applications like complex state classification or fine motor control \cite{duraivel2023high, mathis2024decoding}. Recent works have demonstrated channel counts in the hundreds to thousands \cite{steinmetz2021neuropixels,jun2017fully,sahasrabuddhe2021argo,paulk2022large, musk2019integrated}. In these systems, the AFE is a significant component of the power budget, accounting for 20\% to 99\% of the total \cite{shenoy_paper}. Unfortunately, AFE power scales linearly with the number of simultaneously recorded channels, significantly increasing the power of large recording systems. Developing techniques to reduce AFE power is essential for building implants with high channel counts, which in turn would enable new therapeutic applications.

Decades of circuit design innovations have produced neural‐recording systems whose power consumption now approaches the fundamental limits in the tradeoff between noise and power \cite{chen_trends_of_neural_recording,PrezRecordingStrategies}. The AFE is designed to match the spectral and temporal characteristics of the target neural signals. With further reductions in circuit‐level power increasingly difficult to achieve, the circuits must be evaluated together with the algorithm to find opportunities to save power at the system level.  Recent work has shown that the specifications for AFEs designed for neuroscience research are too stringent, and relaxed design targets can save power while achieving equivalent performance in clinical applications \cite{shenoy_paper}.

Other work has attempted to reduce system power consumption by sub-selecting electrodes from the full array to reduce total power consumption at the cost of accuracy \cite{shin2022neuraltree}, a strategy we refer to as \textit{channel selection}. Neural recordings taken from existing arrays frequently exhibit correlated activity across neighbouring channels \cite{686789,muller2016spatial}, channel selection leverages this redundancy by selecting only those electrodes deemed most informative by the decoding algorithm. However, sub-selecting channels reduces the decoding accuracy while only providing a linear savings in power consumption, limiting the amount of power that can be saved with this approach.

In this work, we propose a technique that records from the full array and reconfigures every AFE resolution at runtime to reduce total power. We call this technique \textit{resolution reconfiguration}. Reducing an AFE's resolution saves power in a greater than linear trade off. For a noise limited design, relaxing resolution by one bit saves a factor of four in power \cite{murmann2008d}. Similar to channel selection, we bring machine learning into the loop by assigning each channel an importance score that represents the effect of that channel on overall classification accuracy. Importance scores are then mapped to resolution settings such that important channels are recorded with high power, high resolution AFEs and less important channels are recorded with low power, low resolution AFEs.


Earlier work has demonstrated that reconfiguring AFE resolution can yield power savings in targeted applications. For example, O’Driscoll et al. showed that analog-to-digital (ADC) resolution can be adjusted on a per-electrode basis in action-potential recording systems \cite{odriscoll_adaptiveADC}. However, because an AFE also contains low-noise amplifiers, filters, and intermediate gain stages, modifying only the ADC resolution reduces just a fraction of the total power consumption. Other studies have found that, in multichannel recording systems, the effective AFE resolution may be relaxed when using probabilistic classification algorithms \cite{galindez_adap1,galindez_adap2,galindez_adap3}. Those evaluations, conducted on human activity recognition datasets, relied on Bayesian decoding to derive channel-specific resolution targets. In contrast to both prior work, our approach assesses the entire signal chain for neural decoding and remains algorithm-agnostic, demonstrating its effectiveness with both linear and nonlinear machine-learning methods.


\textcolor{black}{We compare the simulated performance of a resolution reconfigurable neural recording system with an array of fixed resolution AFEs, which we referred to as a traditional array, for three different recording tasks: 1) seizure detection \cite{cook2013prediction}, \cite{goldberger2000physiobank}; 2) multiclass EMG gesture classification \cite{andy_emg_nature}; and 3) continuous force decoding for a reach and grasp task in non-human primates \cite{grasp_dataset}. Resolution reconfiguration trades off accuracy for power and thus we compare power saving while limiting performance degradation to less than absolute five percentage points from the fixed resolution array. Taking the geometric mean of savings across patients, we find that resolution reconfiguration saves power by a factor of 4.1x, 12.8x and 23.0x when compared to a traditional array for the seizure detection, gesture recognition, and force regression respectively.}

\textcolor{black}{We further compare resolution reconfiguration and channel selection and find that resolution reconfiguration saves power by a factor of 1.6x, 8.4x and 5.6x when compared to channel selection for equivalent performance on the three tasks respectively. The approach also extends to non-linear decoders where resolution reconfiguration saves between 12.5x and 27.2x compared to a traditional array and between 1.9x and 8.2x more than  channel selection for equivalent performance. The spatially distributed tasks of gesture recognition and force decoding show greater savings when compared to channel selection than the localized seizure detection task.}

\textcolor{black}{Lastly, to demonstrate the applicability of this algorithm for on-line decoding, we test resolution reconfiguration in  human participants using forearm surface EMGs to perform a computer cursor control task. With a linear Kalman filter decoder, resolution reconfiguration saved 3.1x compared to a full array and 2.4x compared to channel selection. The off line simulation study and on-line decoding results show resolution reconfiguration to be a system-level tool to save power in neural recording systems. }

\section*{Results}

\subsection*{The Input Referred Noise to Power Consumption Tradeoff}

Figure \ref{fig:figure3} depicts a typical neural‐recording AFE, which comprises a low-noise amplifier (LNA), a variable-gain amplifier (VGA) with integrated low-pass filtering, and an ADC. In power-optimized designs, the LNA and ADC dominate the total consumption, since they establish the thermal-noise and quantization-noise floors. The system’s recording bandwidth, maximum input amplitude, and target signal-to-noise ratio (SNR) together define the minimum power requirements of these blocks. For neuroscience applications, noise from the AFE electronics is engineered to remain well below the level of intrinsic biological background activity.

\textcolor{black}{Although bandwidth and input range specifications both drive AFE power consumption, only the SNR can be adjusted without risking unrecoverable signal loss. The input range requirement is set by the maximum amplitude of neural signals; insufficient range causes the amplifier to saturate, leading to unrecoverable signal loss. Similarly, if the AFE bandwidth is too narrow, out-of-band signal components relevant to decoding are permanently lost. In contrast, lowering the SNR does not induce saturation or signal loss, so decoding algorithms can still operate, albeit with reduced accuracy. We note that bandwidth reconfiguration is a complementary axis of optimization; with appropriate anti-aliasing filtering, bandwidth can be reduced without aliasing, and prior work has shown it can be aggressively reduced with minimal impact on decoder performance \cite{shenoy_paper}. However, this work focuses on SNR scaling, which offers a more favorable power reduction (4x power for every 2x SNR) than the linear power savings from bandwidth scaling.}

\textcolor{black}{Figure \ref{fig:figure3}b depicts the power consumption of an LNA, ADC, and the overall system as a function of SNR for a fixed input amplitude of 10 mV, using the model for power consumption from \cite{shenoy_paper} Since the maximum signal amplitude is fixed by the neural source, this analysis studies the effect of input- referred noise (IRN) as a function of power for the LNA and ADC together. As previously noted, the LNA dominates total system power and delivers a 4x reduction in power consumption for every 2x increase in IRN, when considering white noise sources such as thermal noise and shot noise. This tradeoff does not continue forever due to practical circuit design issues. The analysis ignores flicker noise as flicker noise can be mitigated with chopping and correlated double sampling with little power overhead \cite{denison20072,muller2014minimally,paul2022neural}. Once IRN exceeds roughly  $30 \mu V_{RMS}$, further increases yield little additional power savings since device‐sizing constraints impose a lower bound on power, causing the consumption curve to plateau.}

\subsection*{Increasing Input Referred Noise Reduces Classifier Accuracy}

As noted in prior work, both clinical and research-grade neural recording systems are often over-designed for their target applications \cite{shenoy_paper}. To quantify the effect of IRN on decoding performance, we simulated an AFE whose IRN could be varied, representing various levels of thermal and quantization noise. Since we are simulating a traditional array, all channels have the same resolution and IRN. For each of the three tasks examined in this study, we first trained and tested a linear model on the unaltered data. We then incrementally increased the IRN of all channel AFEs uniformly and measured the resulting degradation in decoding performance (Figure \ref{fig:figure4}) 

For each of the offline decoders, linear models performed well with the original dataset, but a significant amount of noise could also be added without significant degradation in accuracy. For seizure detection from EEG a logistic regression classifier achieved an average $F_1$ score of 0.64 with up to 32 $\mu\text{V}_\text{RMS}$of noise added for a 0.1 loss in $F_1$ score from the Melbourne dataset. For gesture recognition from EMG, a multi-class logistic regression classifier was trained and tested on the original dataset achieving an overall accuracy of 95\%. Up to 2 $\mu\text{V}_\text{RMS}$ could be added without any loss in accuracy. For a force regression task from ECoG, the baseline dataset achieved an $R^2$ of 0.38 using a least-squares decoder and up to 16 $\mu\text{V}_\text{RMS}$ of noise could be added with no loss in accuracy.

Determining the sensitivity of decoding performance to AFE noise allows designers to select IRN targets for the AFE. Since the datasets were recorded with hardware that was over-designed for the application, we apply further power-reduction techniques starting from the inflection point of these curves. We set the resolution of the traditional array such that the AFEs have the largest IRN possible without causing any degradation in performance (marked 5 in Figure \ref{fig:figure4}), and add that amount of noise to the corresponding dataset. In the channel selection paradigm, AFEs are either configured to the highest resolution setting, \boxed{5}, or disabled entirely. In the resolution‐reconfiguration scheme, the highest resolution setting, \boxed{5}, corresponds to the lowest IRN, highest power configuration.

\subsection*{Channel selection can reduce power while maintaining accuracy}

Prior art has shown that for all three datasets, EEG, EMG, and ECoG, decoding can be performed from only a subset of channels. For the CHB-MIT dataset, previous works \cite{khanmohammadi2017adaptive, chung2024single} note that a seizure can be detected from less than 5 channels of the full dataset consisting of 22 channels. In the gesture recognition dataset, the authors note that high prediction accuracy can be maintained with as few as 16 channels \cite{andy_emg_nature}. Similarly, for the force regression task, previous work has shown that on a simpler task, brain state can be classified with just a single channel \cite{valencia2025}, while for similar reach to grasp tasks, high kinematic correlation can be maintained with a much smaller subset of electrodes and features \cite{jang2022, Valencia2023-vh, borra2023}.

To quantify how decoding accuracy degrades as a function of channel count, we recursively drop channels and record the decrease in decoding performance. Beginning with a decoder trained on the full channel set, we iteratively removed the electrode corresponding to the smallest feature weight, then retrained and retested the model, recording the performance on each iteration. This process was repeated until only a single channel remained. After applying this method to each of the three datasets, we found that on average we can drop 74\% of channels for seizure detection, 73\% of channels in gesture recognition, and 77\% of channels in force regression respectively for five percentage point degradation in performance. Because AFE power consumption scales linearly with the number of active channels, these reductions correspond directly to equivalent decreases of 74\%, 73\%, and 77\% in total power usage.

We compare power savings from channel selection at the maximum allowed IRN to power savings from resolution reconfiguration. In most neural recording systems, channel selection can be implemented with little design overhead. The only extra power dissipated by a real system is the leakage power in a channel that is turned off. In contrast, a resolution-reconfigurable design requires a more sophisticated AFE architecture, and thus should achieve substantially greater power savings than channel selection to justify its added complexity.

\subsection*{Resolution reconfigurations consumes less power at equivalent accuracy to channel selection}

\begin{figure}[htp]
    \centering
    \includegraphics[width=\textwidth]{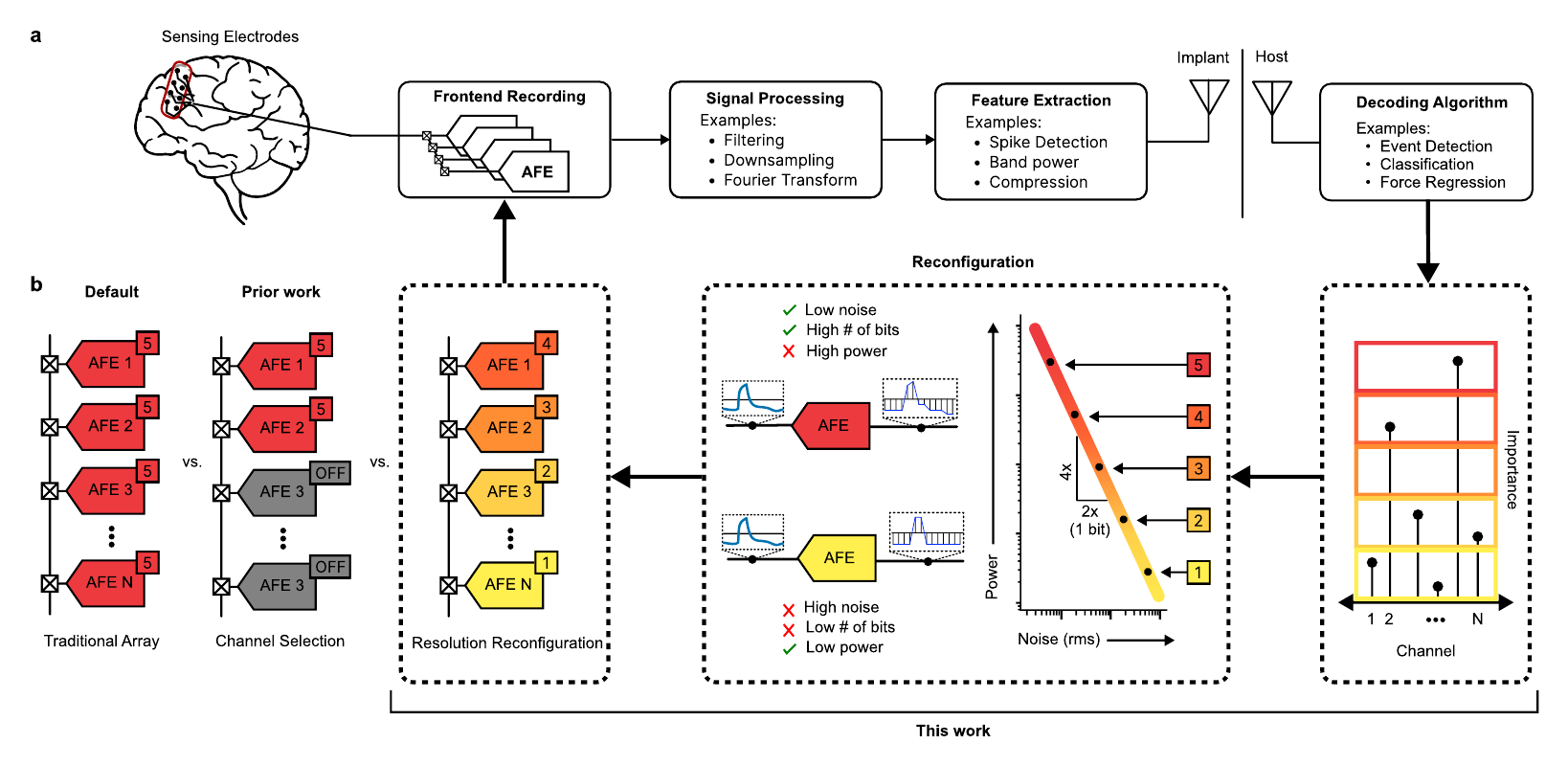}
    \caption{(a) A typical implanted neural recording system. (b) A comparison between three system level approaches for neural recording arrays: 1) a traditional array, 2) a channel selective array 3) a resolution reconfigurable array.}
    \label{fig:figure1}
\end{figure}

To evaluate the power consumption of resolution reconfiguration, we configured the settings of each channel in the following way. During training, all channels were configured to setting \boxed{5} (the highest-performance mode, as shown in Figure \ref{fig:figure4}), identical to the channel-selection benchmark. After training, each channel was assigned a new setting based on the magnitude of its corresponding classifier weight; for multi-feature algorithms, the maximum feature weight defined the channel weight. Both the training and test sets are then re-recorded using the virtual AFE model, which is configured with the assigned settings for each channel. We note that this maps weights that are distributed in a linear scale to SNR which is distributed in log scale, as shown in Fig. \ref{fig:figure1}. 

To maintain decoder performance, it was necessary to retrain the classifier using the new channel settings. Retraining allowed the classifier to incorporate the now higher noise present in specific channels and adjust the model weights to achieve maximum performance. For testing, the channels are configured based on the settings derived from training on the original train set and then evaluated on the rerecorded test set. The full training and test process is outlined in Figure \ref{fig:figure3}. The number of unique settings each channel can have is not fixed. In this work, we evaluate AFEs with up to five unique settings. Each setting is assumed to have one less bit of resolution, corresponding to a 6 dB decrease in SNR, compared to the previous setting. As shown in Figure \ref{fig:figure6}, the array level power consumption can be reduced between 12x and 23x compared to a full array for a five percentage point degradation in performance.

Figure \ref{fig:figure6}e compares the performance of resolution reconfiguration to channel selection. At five settings, resolution reconfiguration saves between 3x and 8x more power than channel selection for equivalent performance. Figure \ref{fig:figure6}f shows that if we limit the degradation to five percentage points, resolution reconfiguration still saves between 3x and 5x more than channel selection. For all three tasks, the proposed resolution reconfiguration performs at least as well as channel selection for equivalent performance in all patients. That is, it always gives higher accuracy for an equivalent power consumption or alternatively consumes less power for a given accuracy target.

\subsection*{Resolution reconfiguration saves power regardless of decoding algorithm}

We further evaluate the performance of the proposed methodology offline on nonlinear decoders. For the seizure detection and gesture classification task, we evaluated a random forest classifier, and for the force regression task, we evaluated the performance of a feedforward neural network (see Methods). For non-linear classifiers that do not have explicit weights we ranked the channels by measuring the degradation in performance when a particular channel is removed. A larger degradation corresponds to a higher importance channel. We then either recursively drop the channel of lowest importance, or use the importance score to map the channel to a resolution setting.

Figure \ref{fig:figure7} compares the performance of channel selection with the performance of resolution reconfiguration. On average, resolution reconfiguration saves between 12x and 27x over a full array. Resolution reconfiguration saves between 2x and 6x more than channel selection for equivalent performance and between 1.25x and 8.25x if we limit degradation to five percentage points. For seizure detection resolution reconfiguration always performs within 1.1x the power over channel selection, while for gesture recognition and force regression, resolution reconfiguration always outperforms channel selection

\subsection*{Power savings are validated through online experiments}

{\color{black}In previous sections, we demonstrate the benefits of resolution reconfiguration in offline contexts. We next evaluate its efficacy during online, closed-loop control. Five participants (n = 5) used wrist movements detected by 5-channel surface electromyography (sEMG) on the forearm to control a virtual cursor in an eight-target center-out target capture task via a linear Kalman filter decoder. For each configuration of settings, where a certain number of channels is dropped or a certain set of resolution settings is used, participants completed a 4-minute visual feedback block. During the visual feedback trials, one visual target appeared on the screen, cueing participants execute the corresponding wrist movements in the absence of cursor feedback. This data is used to train a linear Kalman filter that decodes cursor kinematics, position and velocity, from smoothed root-mean-square (RMS) sEMG features. The trained decoder was then deployed for online cursor control, where the user has real time control of the cursor and aims to move it towards the target. Control performance was quantified using movement time, defined as the total time for the user to reach the target and return to center, capturing the agility of the control. 

Figure \ref{fig:figure8} shows the online results of noise floor sweeping, channel selection, and resolution reconfiguration, respectively. The detailed description of the online experiment protocol is in Method. As seen in Fig. \ref{fig:figure8}d we save an average of 3.1x compared to a full array and 2.4x more than channel selection for a 0.5s degradation in movement time. We note that this study only uses five channels, limiting potential power savings compared to other datasets in this study. Results in Fig. \ref{fig:figure8}c shows that user performance degrades significantly with just one channel, as the task requires control in a minimum of two dimensions to move to all targets. For resolution reconfiguration in the same regime of power savings as channel dropping to a single channel, information from all channels is still present allowing the user to retain the two dimensions of control. This results in resolution reconfiguration performing significantly above channel selection. 

Beyond demonstrating the value of keeping channels with resolution reconfiguration, this task also enables us to compare how our offline results compare with online performance when there is a user in the loop who can adapt to the new settings. We find that when comparing offline to online results, resolution reconfiguration shows greater online gains than channel selection (Supplement: Offline vs. Online Performance in sEMG Cursor-Control), demonstrating that resolution reconfiguration changes information content in a manner that is easier for users to adapt to than channel selection. 
}



\section*{Discussion}

With the growing number of channels in neural recording systems, AFE power consumption is quickly becoming a limiting factor. In this work we evaluated three different system-level approaches to reduce power: adjusting the resolution of the entire AFE array together, channel selection, and channel resolution reconfiguration. The first uses previously published techniques to show that recording specifications for devices used to record the datasets in this work were over-designed for their intended clinical applications \cite{shenoy_paper}. We begin our analysis by reducing the full array resolution as much as possible without incurring any performance loss. Results from both channel selection and resolution reconfiguration demonstrate further power savings beyond the technique proposed in \cite{shenoy_paper}. Further, our results show that resolution reconfiguration is significantly more power efficient than channel selection and saves power compared to a traditional array with fixed resolution. Across both linear and nonlinear decoders, the results show that this method provides power savings comparable to or greater than channel selection, a trend observed across tasks with various channel counts. It can save up to a factor of 27x in power compared to a traditional array, which is 8.2x more than channel selection.

\textcolor{black}{Resolution reconfiguration is more advantageous on problems with high spatial distribution. While savings were consistent between linear and nonlinear decoding algorithms for the gesture recognition and force regression tasks, nonlinear decoding led to little overall power savings on the seizure detection task when compared to channel selection. This may arise from seizure detection being a highly localized problem, improving the efficacy of channel selection. Prior works [41], [42] analyzing the Melbourne dataset demonstrated similar findings, showing that seizure prediction can be performed with a few channels, and our results on the CHB-MIT dataset in the context of detection align with this observation as well (See Supplement). In contrast, for gesture recognition and force regression, the data are distributed across many channels, and even with nonlinear decoding algorithms, channel selection still requires many active channels.}

\textcolor{black}{Resolution reconfiguration also provides a greater advantage over channel selection on lower decoding accuracy users. On the EMG and force regression datasets, resolution reconfiguration provided the greatest efficiency advantage on users or cross-validations that had the lowest overall performance. This may be because, in low accuracy settings, resolution reconfiguration provides more data to the decoder for equivalent power than channel selection, achieving higher accuracy at the same power. Taken together, resolution reconfiguration may provide the greatest advantage over channel selection on subjects or patients that show the lowest overall performance and on applications where the data is distributed across many electrodes.}

\textcolor{black}{For resolution reconfiguration, importance scores must be mapped to resolution settings. In this work, the lowest importance maps to the lowest setting and the highest importance maps to the highest setting. Then, between the maximum and minimum importance score, we construct a linear mapping from importance to setting, binning each channel to a discrete number of settings. This linear map is empirically determined to be effective at saving power while maintaining high accuracy. To evaluate the effectiveness of this mapping strategy, we compared its performance to a random assignment of settings that preserved importance order. The linear map was found to be an efficient assignment scheme (See Supplement). However, there could exist an optimal mapping algorithm that could save even more power, or allow for more configurations that cover a broader space of power and accuracy.}

\subsection*{Outlook}

\textcolor{black}{
We propose a new paradigm for neural-recording AFEs using resolution reconfiguration after deployment. Resolution-reconfigurable AFEs have been implemented for other sensor applications \cite{reconfigurable_sar_adc,reconfigurable_sar_mit, mondal2021dynamically}, and while prior work has studied how power scales with resolution, it made no attempt to optimize this scaling toward the theoretical maximum. While a 4x reduction in power per 2x reduction in resolution represents the theoretical limit, practical circuit design constraints may reduce the achievable scaling factor. However, we show that even scaling factors short of this limit, such as 3x, still deliver significant power savings, including relative to channel selection (See Supplement). Recent work has even demonstrated resolution reconfiguration in a 16-channel system and shown power savings over channel selection for some limited tasks \cite{pandey202636}. Our work shows that as circuits approach this theoretical limit, large power savings can be realized, presenting both a promising opportunity and a new design challenge for the neural-recording circuits community.}

Channel selection and resolution reconfiguration embed machine learning directly into the hardware control loop, but they incur the overhead of computing channel importance scores. This can be computationally expensive for some nonlinear decoders. The exhaustive strategy used here delivers highly accurate importance scores but carries a heavy computational burden that scales quadratically with channel count, making it impractical for arrays with thousands of electrodes. Moreover, recent work has introduced implantable decoders with online update capabilities \cite{chua2022soul,admm_seizure_detector}. Realizing such in situ adaptation would require importance‐scoring routines that are both computationally lightweight and scalable. Improved importance scoring algorithms present a power saving opportunity when combined with resolution reconfiguration from this work. 

\textcolor{black}{While this work contrasts between channel selection and resolution reconfiguration, the two techniques are complementary and could be combined within the same device. Across most of the datasets, we observed that a subset of channels could be removed entirely without any loss in accuracy. A hybrid framework that first excludes uninformative channels and then applies resolution reconfiguration to the active subset could save more total power than either technique alone, further extending the Pareto optimal curve of the power and decoder performance tradeoff. This approach is especially promising for channel-localized tasks such as seizure detection, where a majority of channels can often be excluded without performance degradation.}

\textcolor{black}{The results presented here pertain to a subset of neural‐decoding tasks using arrays of fewer than 100 electrodes, derived from low-frequency biopotentials. We anticipate that, when extended to architectures comprising thousands of channels, this technique may deliver even greater efficiency gains. While highly scaled systems have overhead that is not accounted for in the power budget in this work, we anticipate these overheads to be small relative to the potential power savings. These ideas can also be extended to apply to intracortical spike recording systems that use spike sorting to extract spike trains from a combination of channels \cite{pachitariu2024spike}. This can be achieved by extending the feature weight to channel importance score propagation of this work to apply to non-linear spike sorting algorithms. High channel count spike systems may further benefit from hybrid approaches combining channel selection with resolution reconfiguration due to the localized nature of intracortical spikes when compared to population-level data used in this work. Our findings highlight a compelling opportunity to co‐optimize circuit design with decoding algorithms, enabling neural‐recording systems to scale by an order of magnitude in channel count without increasing their power footprint.}

\section*{Acknowledgments}
We would like to thank the Semiconductor Research Corporation, JUMP 2.0, Center for Codesign of Cognitive Systems (CoCoSys) award AWD-004311-S2, the NSF Graduate Fellowship, the Apple fellowship in Integrated Systems, Weill Neurohub Next Great Ideas Award, and the sponsors of the Berkeley Wireless Research Center. {\color{black}We also thank Dr. Dean Freestone and Dr. Mark Cook for providing the intracranial electroencephalography (iEEG) patient dataset.}

\section*{Competing Interests}

{\color{black}R.M. receives consulting fees from Synchron, Inc., a company operating in a field related to the subject matter of this manuscript. Synchron, Inc. was not involved in the design, execution, analysis, interpretation, or funding of this work. A.P. was employed at Analog Devices, Inc a company operating in a related field to the subject matter of this manuscript following initial submission of this manuscript. Analog Devices was not involved in the design, execution, analysis, interpretation, or funding of this work. } 

\section*{Author Contributions}

{\color{black}A.P. and R.M. conceived of the idea and designed the experiments. I.L. conducted experiments on the seizure detection dataset, A.P. conducted experiments on the gesture recognition dataset, and D.V. conducted experiments on the force regression dataset. A.P., D.V., and I.L. analyzed and compiled the results. H.S., I.L., and D.V. conducted the online experiment, and I.L. processed the data. P.K. provided the materials, protocol, and IRB for the experiment. A.P. wrote the original manuscript and R.M. revised the manuscript. A.P. designed the figures and D.V. revised them. R.M. oversaw the project. All authors reviewed and approved the final manuscript.} 

\section*{Data Availability}

{\color{black}The sEMG gesture classification dataset, force decoding dataset, and the CHB-MIT dataset are publicly available and can be accessed through their respective publication. The Melbourne iEEG dataset contains confidential participant information and is available under a Creative Commons license with restrictions on ATTRIBUTION, NON-COMMERICIAL use, (DOI: \url{https://doi.org/10.26188/5b6a999fa2316}). The Melbourne iEEG dataset access should be requested from \url{https://www.epilepsyecosystem.org/howitworks/#data} ("Melbourne Seizure Prediction Trial Seizure Data"). Users will be required to create an account and sign a terms of use agreement that requires no commercial use, and restricts all works derived from the data to be made publicly available under the same license. This work used a subset of the Melbourne iEEG dataset received from Dr Mark Cook. The sEMG online cursor control data is available at \url{https://github.com/MullerGroup/resolution\_reconfigurable\_modeling.git}.}  

\section*{Code Availability}

The code developed for this study is available at
\url{https://github.com/MullerGroup/resolution\_reconfigurable\_modeling.git.}
\clearpage

%

\clearpage

\begin{figure}[htp]
    \centering
    \includegraphics[width=\textwidth]{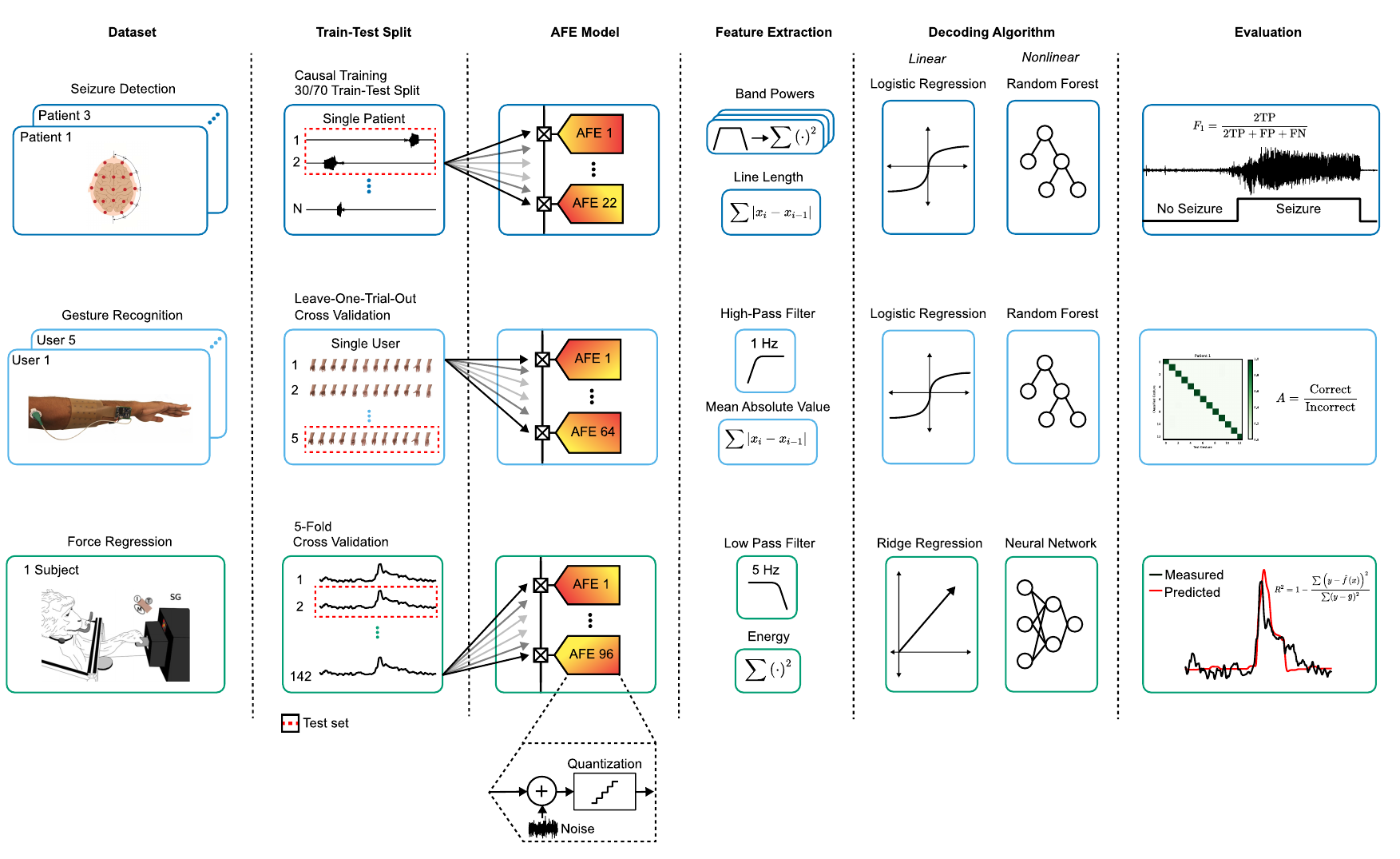}
    \caption{Schematic diagrams of the processing steps used for each of the three different datasets used in this work. (a) The seizure dataset, (b) The EMG gesture recognition dataset, and (c) the force regression dataset.}
    \label{fig:figure2}
\end{figure}


\clearpage

\begin{figure}[htp]
    \centering
    \includegraphics[width=\textwidth]{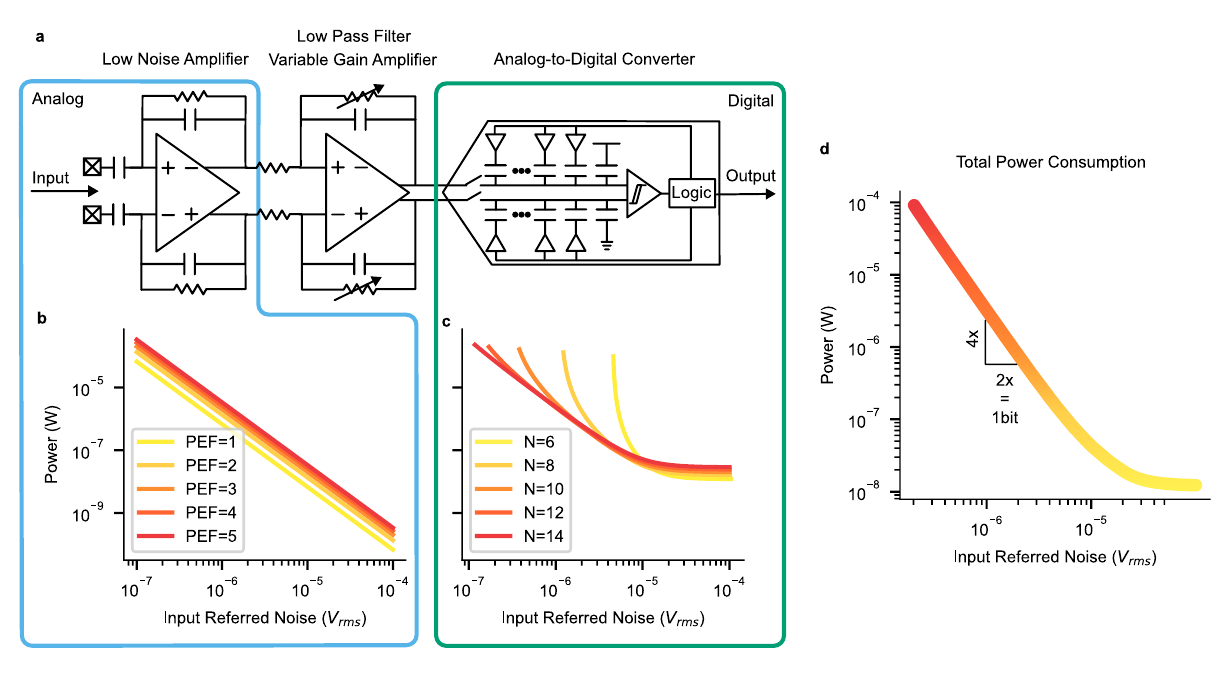}
    \caption{(a) A common neural signal recording AFE consisting of a low noise amplifier, filter, variable gain amplifier, and Successive Approximation Register (SAR) based ADC. (b) Low noise amplifier power consumption vs input referred noise at different Power Efficiency factors (PEF). (c) SAR ADC power consumption as a function of the input referred noise at various ADC resolutions (N). (d) Power consumption of the entire AFE as a function of the input referred noise (See Supplement).}
    \label{fig:figure3}
\end{figure}


\clearpage

\begin{figure}[htp]
    \centering
    \includegraphics[width=\textwidth]{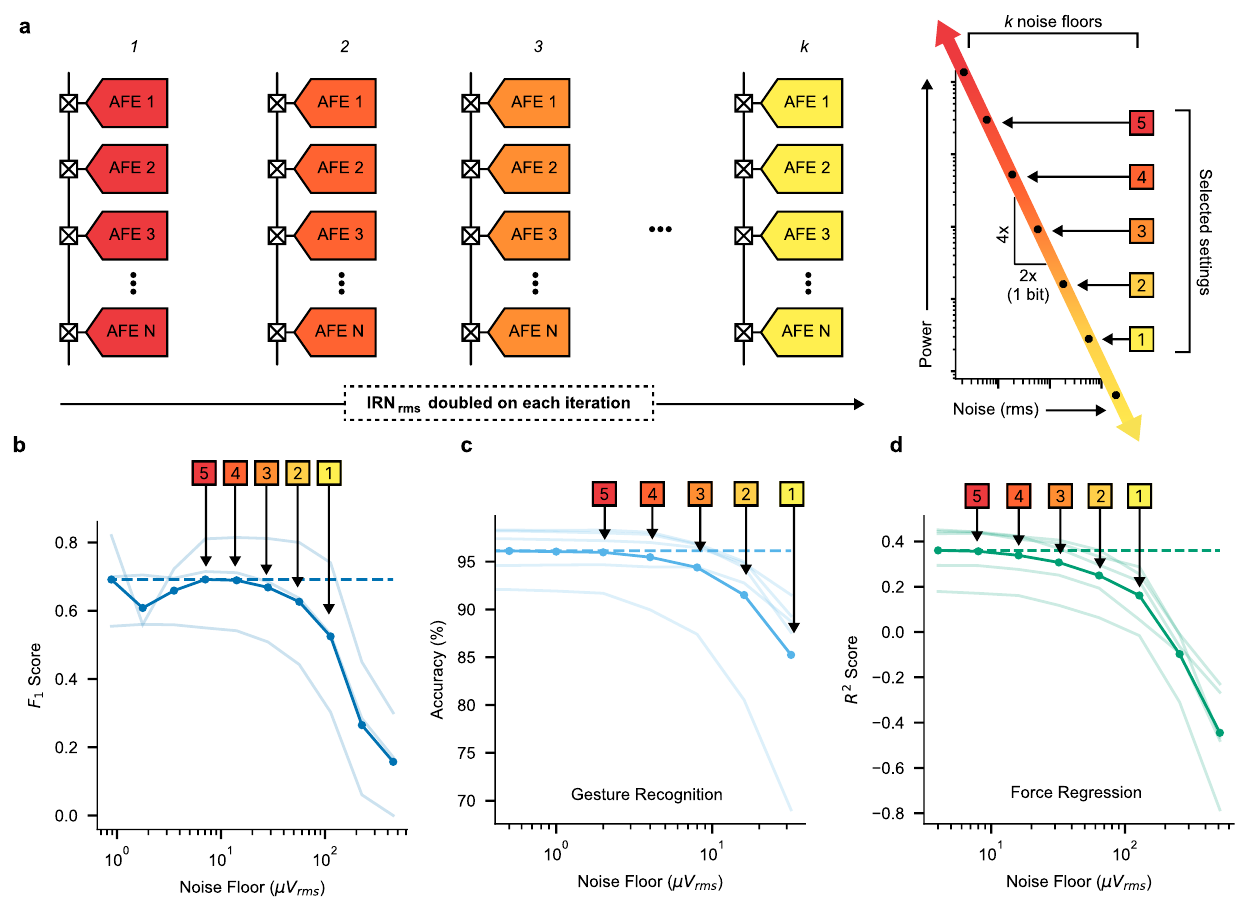}
    \caption{Performance results of changing the input referred noise (IRN) of every AFE together in each dataset. (a) Shows the configuration of every AFE across the sweep. {\color{black}(b) The $F_1$-score of a logistic regression classifier on the seizure dataset as a function of IRN.} (c) The accuracy of a multi-class logistic regression classifier on the gesture dataset. (d) $R^2$ of a linear regressor on the motor decoding dataset. Lines represent every patient or cross validation in the dataset (see Methods). Boxed numbers represent the selected IRN levels for each dataset, \boxed{5} is always selected to be the highest IRN that leads to no loss in accuracy.}
    \label{fig:figure4}
\end{figure}


\clearpage

\begin{figure}[htp]
    \centering
    \includegraphics[width=\textwidth]{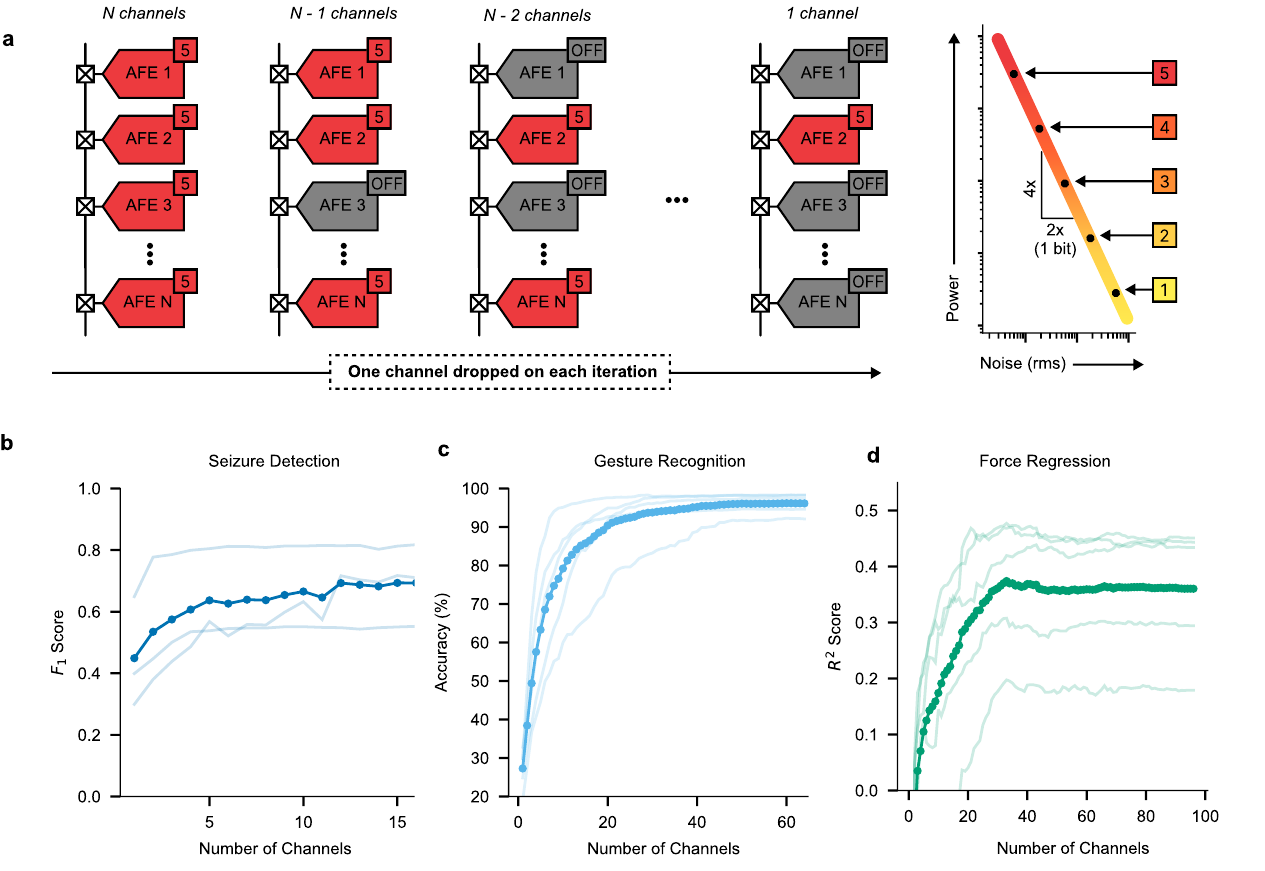}
    \caption{Performance results with channel selection for each dataset. (a) Shows an example configuration of the AFEs for this simulation. At every iteration, the AFE channel with the smallest overall importance is removed and the classifier is trained and tested again (see Methods).  Sub-figures{\color{black}(b)}(c)(d) show the results of channel selection for the seizure detection, gesture classification, and motor decoding datasets respectively.}
    \label{fig:figure5}
\end{figure}


\clearpage

\begin{figure}[htp]
    \centering
    \includegraphics[width=\textwidth]{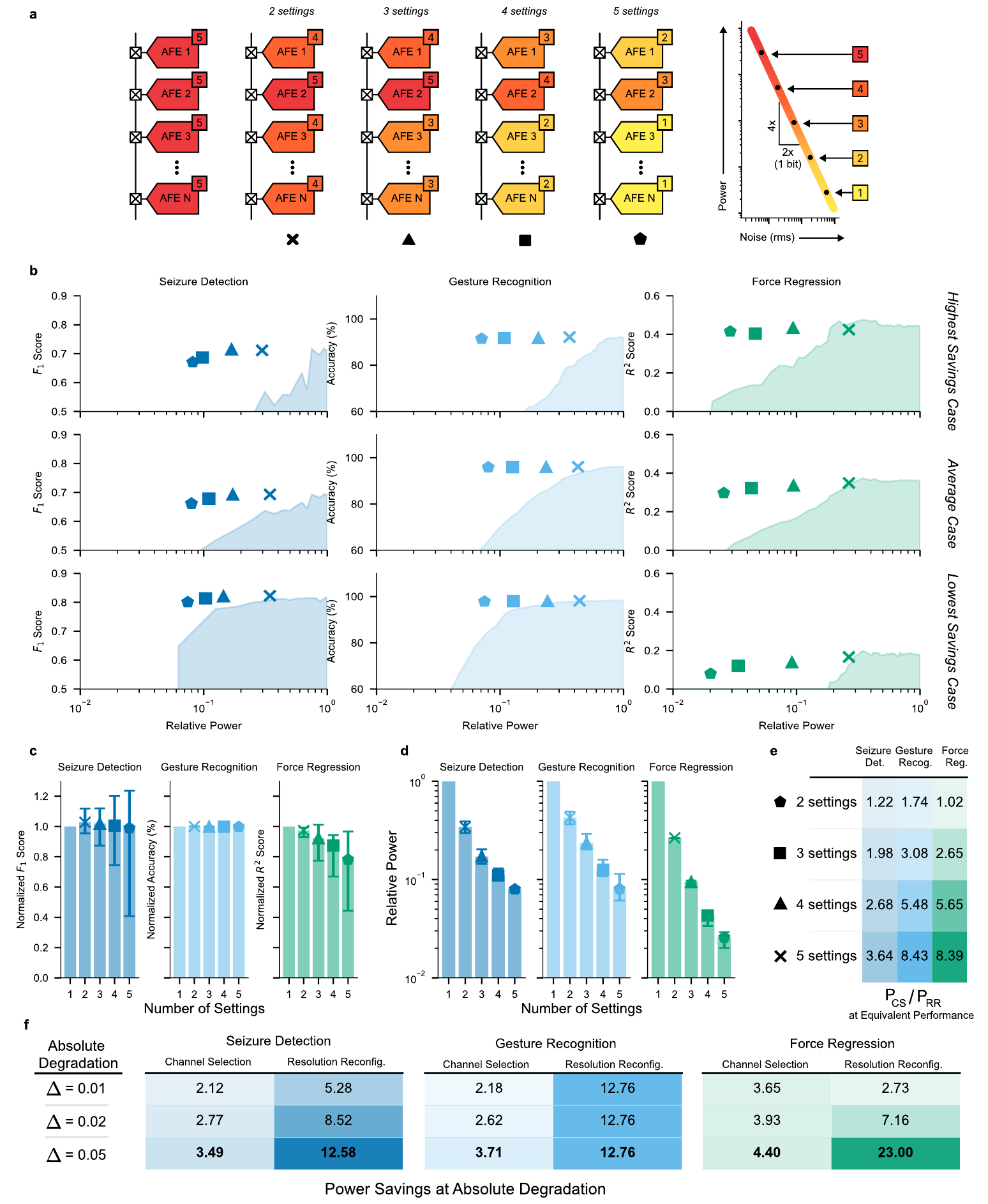}
    \caption{{\color{black}(a) Shows an example configuration of the AFEs in each of the different channel reconfiguration scenarios from a system with 2 unique settings to a system with 5 unique settings. (b) Shows a comparison between the channel reconfiguration technique (shown with different points) and the channel selection technique (boundary of shaded region) for the best case, average case, and worst case for each dataset. Shaded regions represent areas that are less efficient, that is more power for less accuracy than channel selection. (c) Shows the relative accuracy metric corresponding to each classification problem across the different configurations. Error bars are shown for cases where there is more than one setting. (d) Shows the relative array level power consumption of the AFEs across the different settings. (e) Shows a table containing the ratio of channel selection power to resolution reconfiguration at the same accuracy as achieved by resolution reconfiguration. Higher ratios imply resolution reconfiguration saves more power. (f) Shows tables comparing resolution reconfiguration and channel selection power savings compared to a traditional array for a given absolute accuracy metric degradation target.}}
    \label{fig:figure6}
\end{figure}


\clearpage

\begin{figure}[htp]
    \centering
    \includegraphics[width=\textwidth]{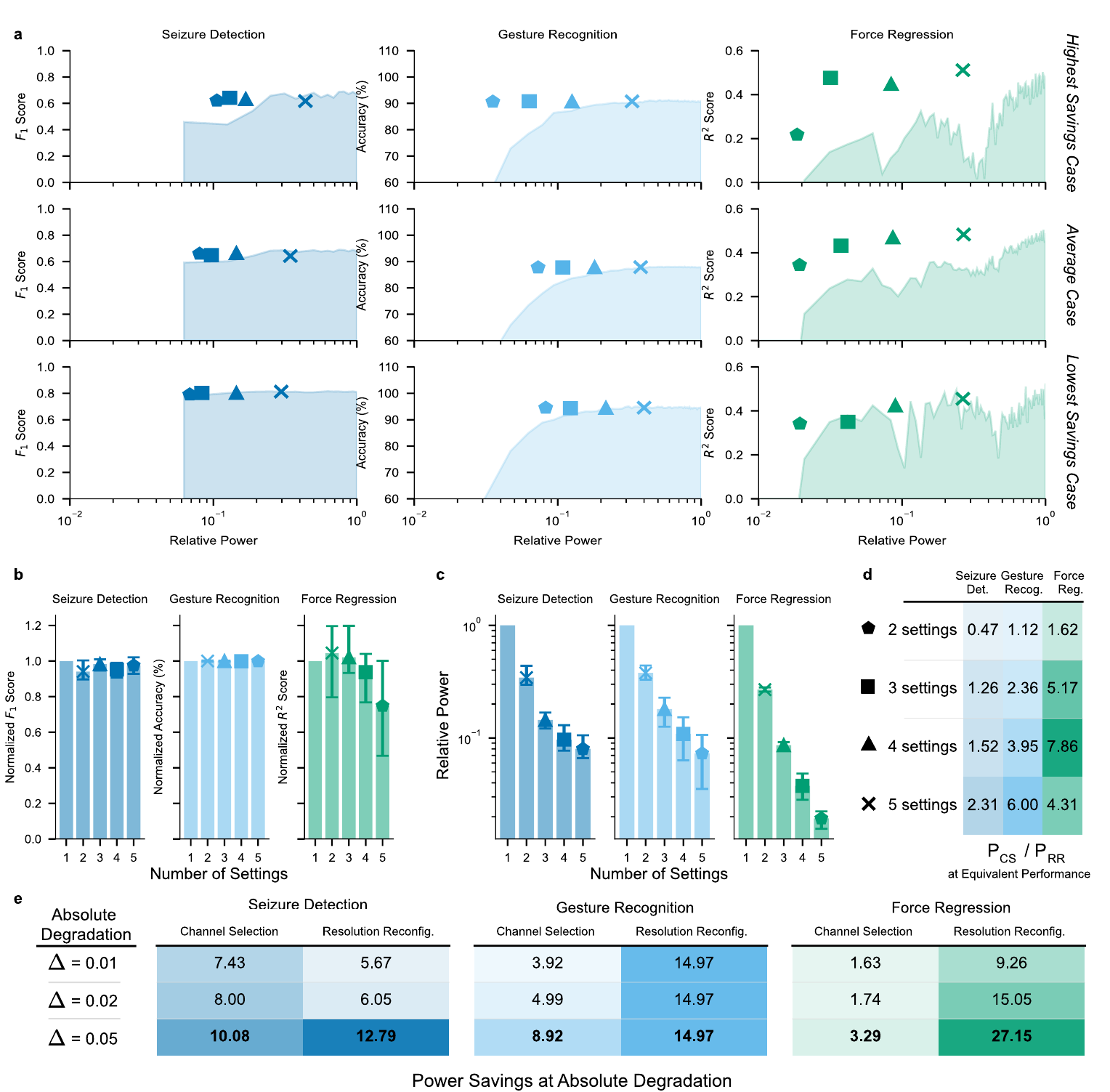}
    \caption{{\color{black}This figure shows the results of the same experiment as shown in Figure 6, but using a non-linear model for classification. (a) Shows a comparison between the channel reconfiguration technique (shown with different points) and the channel selection technique (boundary of shaded region) for the best case, average case, and worst case for each dataset. Shaded regions represent areas that are less efficient, that is more power for less accuracy than channel selection. (b) Shows the relative accuracy metric corresponding to each classification problem across the different configurations. Error bars are shown for cases where there is more than one setting. (c) Shows the relative array level power consumption of the AFEs across the different settings. (d) Shows a table containing the ratio of channel selection power to resolution reconfiguration at the same accuracy as achieved by resolution reconfiguration. Higher ratios imply resolution reconfiguration saves more power. (e) Shows tables comparing resolution reconfiguration and channel selection power savings compared to a traditional array for a given absolute accuracy metric degradation target.}}
    \label{fig:figure7}
\end{figure}


\clearpage

\begin{figure}[htp]
    \centering
    \includegraphics[width=\textwidth]{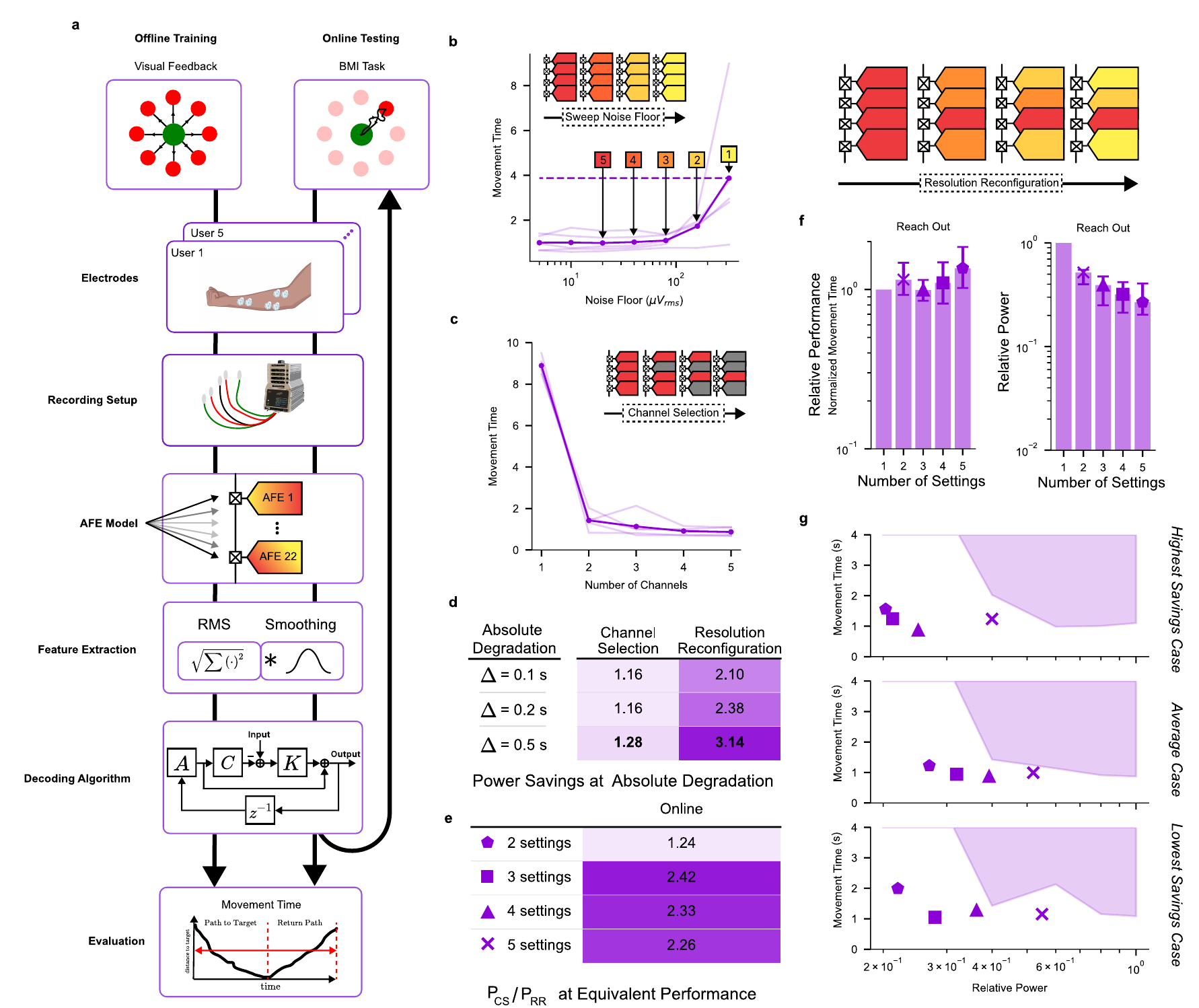}
    \caption{{\color{black}Online sEMG results. (a)-(c) show the relative AFE power as a function of the number of settings, across different AFE power scaling factors. (d)-(f) compare the channel reconfiguration power and sparse channel selection power at equivalent accuracies as a function of the number of settings across different AFE power scaling factors.}}
    \label{fig:figure8}
\end{figure}


\clearpage
\section*{Methods}

\subsection*{Noise Modeling}

For each of the datasets, the recorded data is considered true voltages, including any noise sampled by the original data acquisition system. For each AFE setting, two forms of noise are added to model a lower performance system, thermal noise and quantization noise. The thermal noise is assumed to be white, that is, it has a uniform distribution across frequency. Quantization noise is added by rounding the voltages to equivalent levels. {\color{black}The quantization levels are set such that the quantization noise is 12dB below the thermal noise. This ratio is accepted to be an optimal design point for low noise, high resolution AFEs \cite{sigma_delta_book}.}

\subsection*{Seizure Detection}

{\color{black}The intracranial EEG (iEEG) dataset collected from a study conducted by the University of Melbourne \cite{cook2013prediction} is used to evaluate our approach on seizure detection. It consists of three patients that had the lowest seizure prediction performances out of the ten patients in a clinical trial done in \cite{cook2013prediction}. Each EEG recording has 16 channels of data sampled at 400 Hz. In this dataset the recording length of each patient is $>$100 hr, with abundant seizure events. The seizure onsets are defined based on clinician-provided labels.

The first 30\% of the recording of each patient is used as training data, while the remaining 70\% is used as testing data. This causal data splitting scheme better simulates an online deployment scenario.

A total of eight features are extracted from EEG signals in 1-second windows to train the decoder. Six spectral band powers—delta (0.1–4 Hz), theta (4–8 Hz), alpha (8–12 Hz), beta (12–30 Hz), gamma (30–70 Hz), and high gamma (70–125 Hz)—are computed using a 6th-order infinite-impulse-response (IIR) Butterworth filter. The remaining two features are line length \cite{esteller2001line} and the total spectral power of the signal. To enable the decoders to discriminate seizure events during periods of weak signal amplitude, the six band powers are normalized by the total spectral power of the signal, and the line length is normalized by the square root of the total spectral power.

During testing, the output seizure probability is smoothed using a 5-tap moving average filter for stability. A seizure is detected in a given window if the output probability exceeds a threshold of 0.5.
The $F_1$ score is used as the decoder performance metric, as it provides a harmonic mean of recall (sensitivity) and precision. This metric is particularly useful for evaluating performance on imbalanced datasets \cite{sokolova2009systematic}, such as the Melbourne dataset. The $F_1$ score is evaluated as:
\begin{equation}
F_1 = \frac{2\times TP}{2\times TP + FN + FP}
\end{equation}
where $TP$, $FP$, and $FN$ denote the number of true positives, false positives, and false negatives within 1-second detection windows, respectively. 
}

\subsection*{Gesture Classification}

Gesture classification was performed on an EMG dataset collected in \cite{andy_emg_nature}. The dataset consists of surface EMG recordings from 64 sites on 5 different users. Each user performs 11 gestures for 5 trials, leading to 55 total trials per user. Each trial is 11 seconds long and consists of a rest period, a transition period, and a gesture hold period, followed by another transition period and a return to rest. Before classification, the rest and transition periods were removed from the data.

Before performing classification, the data was high-pass filtered at 1 Hz, and the mean absolute value was computed for 50 ms windows. The mean absolute value was used as the only feature corresponding to each channel. Data was normalized to mean 0 and standard deviation 1, and then classified through a multi-class classifier that determined the current gesture from the 11 gesture dataset.

For each patient, 1 trial was held out during classifier training and then used to test the classifier. This was then repeated across each of the five trials leading to a leave-one-trial-out cross validation scheme. For each trial, the 11 gestures were compared to each other and overall accuracies were determined. The accuracy corresponding to one patient is the average of the accuracy across all cross validations. 

\subsection*{Continuous Decoding}

To demonstrate our method on a continuous decoding problem, we performed regression on 96-channel ECoG data for continuous force decoding in rhesus macaques. The dataset from \cite{brochier_massively_2018}, consists of 96-channel intracortical LFP from Monkey L and grip force data from four sides of a reaching target. For our study, we extract the 142 correct trials and regress on time series data from the entire duration of the trial. Each trial is approximately 3.5 seconds long. The output regression signal is the sum of force from each of the four sides of the cube. 

In this study, we tested a Ridge regression and a Feedforward Neural Network model on this task. To validate, we use 5-fold group cross validation where each trial is a group to avoid leakage of test data into the training set. $R^2$ correlation coefficient is our performance metric. Since the data is approximately $\frac{2}{3}$ inactive time and $\frac{1}{3}$ active time, the overall $R^2$ is very good but may not accurately represent the actual force being captured. To measure our ability to regress on the true force, we train on the samples from the entire trace, but only compute the $R^2$ over the active portion of the waveform, determined by a threshold set to the mean of the signal. This amounts to fine grained decoding of the actual applied force between high force and low force trials. 

Previous studies have shown that the local motor potential, an evoked low frequency ($\leq$5 Hz) potential present in the motor cortex during the onset of movement, is effective at predicting the onset of movement \cite{flint_long_2013}. To preprocess the data, we extract out the local motor potential (LMP) using a second-order Butterworth filter with cutoff frequency set to 3 Hz. Since we know that the bandwidth of the movement is low ($\leq$1 Hz), we also filter the output data to 3 Hz to improve regression smoothness. The final feature vector consists of a single sample in time of the filtered signal across 96 channels $\in \mathbb{R}^{96}$. The output is a single scalar representing a single sample in time of the smoothed total force signal. While the literature shows that the high frequency gamma and high gamma bands are also helpful for decoding, extensive feature examination showed that these bands did not contribute significantly to improving our metrics and feature selection algorithms consistently showed that the LMP was the strongest covariate. We only tune hyperparameters once, in the initial fit to keep model parameters and learning ability consistent across studies. 

\subsection*{Importance determination}

Both the sparse channel selection algorithm and the AFE setting reconfiguration algorithm require an importance determination step. In this step, the relative importance of each channel in the dataset is determined. For linear decoding algorithms such as logistic regression or ridge-regression, the importance is absolute value of the weight determined by each algorithm. For datasets that computed multiple features per channel, the maximum absolute value is considered the importance of the entire channel. The seizure detection and gesture recognition datasets use random forest classifiers as  a non-linear classifier. The impurity based importance score was used to determine channel importance. The maximum score was taken for classification problems with multiple features per channel. For the neural network classifier, the importance is determined by finding the sensitivity of the classification accuracy metric ($R^2$) to each channel. That is, the accuracy of the classifier with and without the presence of a channel is compared to find the relative importance of each channel.

\subsection*{Sparse Channel Selection}

To find how many channels can be removed from the dataset and in what order, we used a recursive channel removal algorithm. At every step, the classification algorithm was trained and the importance of each channel was determined. The channel with the smallest importance was removed for the next iteration and the process is continued until only one channel remains. For {\color{black}EMG dataset}, this process is performed for each cross validation, and the accuracies are averaged together at the end.

\subsection*{AFE Noise Adjustment}

{\color{black}To adjust the AFE noise levels and find how much power can be saved, we began by training the selected classifier on the dataset and determining the relative importance of each channel.} Then, the relative importances were mapped linearly to the log of the input referred noise. This was used to select the setting for each channel, after which the training data was re-recorded by adding the corresponding amount of noise. The decoder was then retrained on this noisy training data before finding the final weights for testing the decoder.

\subsection*{Online sEMG Cursor Control}

\textcolor{black}{All procedures were approved by the institutional review board at University of California, Berkeley (CPHS \# 2024-02-17167), and informed consent was obtained from all participants. Five participants completed myoelectric control of a computer cursor using surface electromyography (sEMG). Participants did an eight-target center-out target capture task, moving their computer cursor from the center of the screen to a peripheral target and then back to the center. Five commercial Ag/AgCl wet electrodes (Duotrode™, Myotronics, USA) were positioned on each participant's forearm targeting major muscle groups: pronator teres (pair 1), extensor pollicis longus/brevis (pair 2), flexor carpi (pair 3), extensor carpi (pair 4), and extensor digitorum (pair 5). Signals from the sEMG electrodes were amplified and recorded via a Tucker-Davis Technologies S-Box, Subject Interface preamp, and then RZ2 BioAmp processor. To emulate different analog front-end (AFE) recording resolutions, the Real-Time Processor Visual Design Studio (RPvdsEX) was used to inject pseudo-random Gaussian noise at varying levels into each channel post-digitization and prior to featurization. Features were extracted in Synapse software using a pipeline that included a bandpass filter between 10Hz and 100Hz, a notch filter at 60 Hz and 120 Hz, and finally a calculation of a low pass filtered (cutoff is 3 Hz) root-mean-square (RMS) feature on each channel. These features were then streamed to a separate  computer executing cursor decoding and task control.}

\textcolor{black}{After initial training to get familiar with the task, participants completed experimental blocks. Each block began with 4-minutes of open-loop visual feedback where one visual target appeared on the screen, cueing participants to execute the corresponding wrist movement in the absence of cursor feedback. sEMG features extracted from this phase were used to train a linear Kalman filter to decode cursor position and velocity. This fixed decoder was then deployed for a 5-minute closed-loop cursor control phase.}

\textcolor{black}{To establish the baseline noise floor for channel selection and resolution reconfiguration, the injected Gaussian noise was swept in a manner similar to the offline datasets. Starting at 5 $\mu$Vrms, the noise floor was doubled in successive experimental blocks up to a maximum of 320 $\mu$Vrms. The movement time required to reach the target at each level was recorded (Fig. \ref{fig:figure8}). The baseline was fixed to 20 $\mu$Vrms based on the initial characterization of the first participant. This specific threshold was selected to ensure consistent performance across all subjects and to standardize the protocol for subsequent experiments.}

\textcolor{black}{To establish the parameters for subsequent resolution reconfiguration and channel selection, each participant started with a baseline block utilizing all five channels with 20 $\mu$Vrms added noise. Channel importance was assessed offline using the 5-minute closed-loop data from this baseline block. For each channel, its data was removed, the Kalman filter was then retrained, and the resulting cursor trajectory was simulated.
The importance of a channel was quantified by the degradation of a trajectory metric. Specifically, “deviation to target,” is used. It is defined as the average perpendicular distance between the simulated cursor trajectory and an ideal straight line from start to target. This spatial metric over movement time allowed for comparison between online and offline results. Because offline simulations rely on temporally fixed recorded data, they cannot accurately capture dynamic, closed-loop changes in movement time. To determine the channel selection order, the least important channel was iteratively removed until only one channel remained similar to the offline study. For resolution reconfiguration, the calculated channel importance scores were normalized to a 0–1 range. This range was linearly quantized based on the number of available settings. These selection orders and noise configurations were then applied to all subsequent experimental blocks. All subsequent experimental blocks followed the identical procedure of open-loop training and closed-loop control. To evaluate cursor performance during these modified closed-loop phases, movement time is selected as the primary metric, as it is commonly used to quantify target capture performance and encapsulates multiple aspects of control like speed of movement, directness of movement toward the target, and dexterity in capturing the target with fine movement.}

\printbibliography

\end{document}


\maketitle

\section*{More details on Power Modeling}

To model the power consumption of an AFE, we assume the AFE consists of three stages, a Low Noise Amplifier (LNA), a filtering gain stage, and a Successive Approximation Register-based Analog to Digital converter (ADC). In real designs, some of these elements may be multiplexed between channels to save power, however we ignore these cases for this analysis. To model the power consumption of a Low Noise amplifier, we assume the amplifier achieves a given Power Efficiency Factor (PEF) \cite{muller_0013_2012}:

\begin{align*}
\text{PEF} &= \text{NEF}^2V_{DD} = v_{n,i,rms}^2\times\frac{2I_DV_{DD}}{\pi kTf_{bw}} \\
P_{amp} &= V_{DD}I_D = \text{PEF}\times\frac{\pi kTf_{bw}}{2v_{n,i,rms}^2}
\end{align*}

The PEF represents how power efficient an amplifier topology is at achieving a target input-referred noise. Figure 3 shows the power consumption as a function of input referred noise at different amplifier PEFs. For further analysis, we assume the amplifier achieves a PEF of 1, which is among the highest reported PEFs \cite{pef_1_amp1, pef_1_amp_2}.

To model the power consumption of the ADC, we use the analysis from \cite{shenoy_paper}:

\begin{align*}
    C_\text{unit} &= \text{max}\left(C_\text{min}, 24kT\times\frac{snr}{2^\text{nbits}v_{in,p2p}^2}\right) \\
    E_\text{DAC} &= \sum_{i=1}^{\text{nbits}-1}2^{\text{nbits-3-2*i}}(2^i-1)C_\text{unit}V_{ref}^2 \\
    E_\text{COMP} &= \text{max}\left(C_{min}, 24kT\frac{snr}{v_{in,p2p}^2}\right) \\
    E_\text{logic} &= E_\text{gate}\times\text{nbits} \\
    P_\text{ADC} &= 2f_{bw}(E_\text{DAC}+E_\text{COMP}+E_\text{logic})
\end{align*}

Here, we assume $C_\text{min}$, which represents the minimum capacitance that can be reliability fabricated in the process is $1$fF. Further, we assume the energy of a gate required in the SAR ADC is $1$pJ. 

The final component of the power consumption is the filtering gain stage. We assume the filtering gain stage is limited by the required bandwidth, not the noise, as the low noise amplifier has already amplified the signal. As such, assuming a single stage amplifier, we can compute the required current assuming a transconductance efficiency $\frac{gm}{I_d}$. For this work, the transconductance efficiency was assumed to be 20.

\begin{align*}
    P_\text{AMP} &= V_{DD}\frac{g_m}{\frac{g_m}{I_d}} \\
    g_m &= A_v(2f_{bw}n_\tau)C_\text{UNIT}2^\text{nbits} \\
    n_\tau &= -\text{Log}\left(\frac{1}{2^{\text{nbits}+1}}\right)
\end{align*}

\newpage
\section*{Seizure Detection Results from CHB-MIT Dataset}

\begin{figure}[htp]
    \centering
    \includegraphics[width=1.0\linewidth]{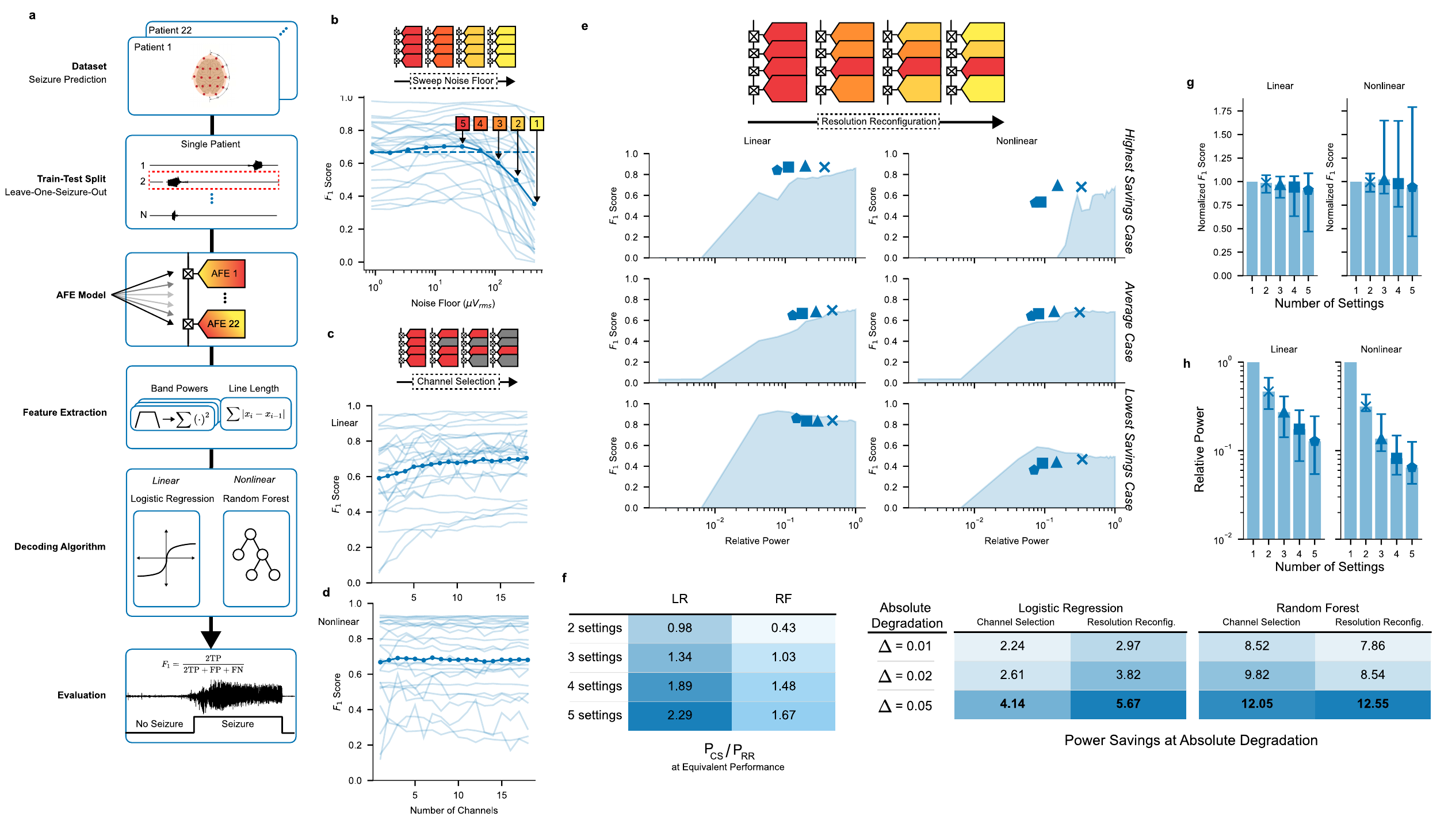}
    \caption{{\color{black}Results from CHB-MIT dataset. (a) illustrates the simulation pipeline. (b) shows the $F_{1}$-score of a logistic regression classifier as a function of IRN. (c) and (d) shows the channel selection for linear logistic regression classifier and nonlinear random forest classifier, respectively. (e) demonstrates the comparison between the resolution reconfiguration (shown with different points) and channel selection (boundary of shaded region) for the highest, average, and worst power saving scenarios. (f) lists the power ratio between channel selection and resolution reconfiguration at equivalent decoder performance across resolution reconfiguration settings. It also shows the power savings from channel selection and resolution reconfiguration for three absolute performance degradation thresholds. (g) shows the normalized $F_{1}$-score across resolution reconfiguration settings for linear and nonlinear classifiers. (h) shows the relative array level power consumption of the AFEs across different settings for linear and nonlinear classifiers.}}
    \label{fig:CHBMIT}
\end{figure}

{\color{black}The Children's Hospital Boston and the Massachusetts Institute of Technology (CHB-MIT) scalp EEG dataset \cite{goldberger2000physiobank}, consisting of EEG recordings from 24 patients, is an additional dataset used to evaluate our approach on seizure detection. Each EEG recording is sampled at 256 Hz with 16-bit resolution. For most patients, 22 EEG channels are used to train the decoder, except in cases where electrode placements are inconsistent across recordings.

Due to the rarity of seizure events, we employ a leave-one-recording-out cross validation scheme to assess decoder performance. In each iteration, one seizure-containing recording is designated as test data, while the remaining recordings serve as training data. The decoder's performance is averaged over all possible training-testing configurations.

A total of eight features are extracted from EEG signals in 1-second windows to train the decoder. Six spectral band powers -- delta (0.1–4 Hz), theta (4–8 Hz), alpha (8–12 Hz), beta (12–30 Hz), gamma (30–70 Hz), and high gamma (70–125 Hz)—are computed using a 6th-order infinite-impulse-response (IIR) Butterworth filter. The remaining two features are line length \cite{esteller2001line} and the total spectral power of the signal. To enable the decoders to discriminate seizure events during periods of weak signal amplitude, the six band powers are normalized by the total spectral power of the signal, and the line length is normalized by the square root of the total spectral power.

During testing, the output seizure probability is smoothed using a 5-tap moving average filter for stability. A seizure is detected in a given window if the output probability exceeds a threshold of 0.5.
The $F_1$ score is used as the decoder performance metric, as it provides a harmonic mean of recall (sensitivity) and precision. The $F_1$ score is evaluated as:
\begin{equation}
F_1 = \frac{2\times TP}{2\times TP + FN + FP}
\end{equation}
where $TP$, $FP$, and $FN$ denote the number of true positives, false positives, and false negatives within 1-second detection windows, respectively. 

Figure \ref{fig:CHBMIT} demonstrates the results from the CHB-MIT dataset. We apply the same process of input-referred noise sweeping, channel selection, and resolution reconfiguration (see Methods) to both a linear logistic regression classifier and nonlinear random forest classifier. The logistic regression classifier achieved an average F1-score of 0.67 when no noise is added to the original dataset. As we sweep the injected input-referred noise floor, the logistic regression classifier can tolerate up to 28 $\mu V_{RMS}$ of noise without losing any decoder performance. This input-referred noise floor (the inflection point in Fig. \ref{fig:CHBMIT}b, marked as \boxed{5}) is chosen as the baseline added noise in the subsequent analysis for logistic regression classifier. Fig. \ref{fig:CHBMIT}(c)-(d) show the results of channel selection for logistic regression classifier and random forest classifier, respectively. We found that on average we can drop 77$\%$ and 92$\%$ of channels for five percentage point degradation in performance for logistic regression and random forest classifier, respectively. Fig. \ref{fig:CHBMIT}(e) shows the results of resolution reconfiguration with different number of settings compared to channel selection for linear and nonlinear classifiers. At five settings, resolution reconfiguration saves 2.29x and 1.67x more power than channel selection for equivalent performance.

We note that resolution reconfiguration led to little overall power savings on CHB-MIT dataset when compared to channel selection, especially with nonlinear decoder. This may arise from seizure detection being a highly localized problem, improving the efficacy of channel selection. The CHB-MIT dataset is known to contain at least three types of seizures: partial, complex partial, and generalized tonic-clonic \cite{goldberger2000physiobank}, \cite{ali2024epileptic}. Partial and complex seizures are localized by definition \cite{fisher2017operational}. Our results from channel selection showed that accurate seizure detection can be performed on as few as one channel for some users when using nonlinear decoding algorithms. This is aligned with the finding from prior work \cite{chung2024single}, which shows that neurologists can identify the distinct, localized channel over half of the patients in the CHB-MIT dataset. Notably, in some of the patients the decoder’s performance is not monotonically decreasing when the number of channels is reduced. This is also suggested by \cite{chung2024single} that removing channels could reduce the artifacts or noise injected from irrelevant channels, thereby improving the model testing accuracy.}

\clearpage
\section*{Offline vs. Online Performance in sEMG Cursor-Control}

\begin{figure}[htp]
    \centering
    \includegraphics[width=1.0\linewidth]{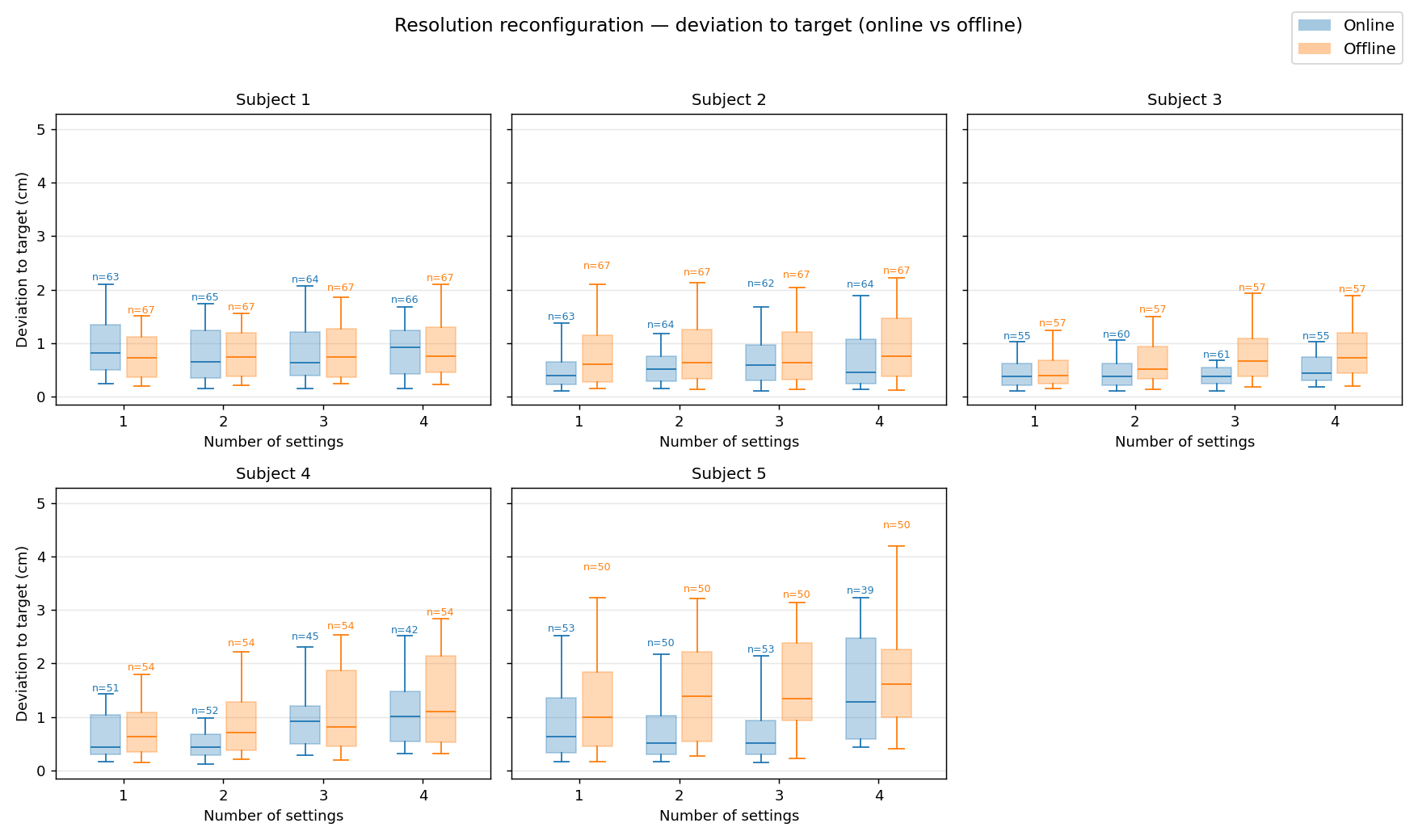}
    \caption{{\color{black}Online decoder vs. Offline decoder performance of five subjects given same resolution reconfiguration settings.}}
    \label{fig:rr_offline_vs_online}
\end{figure}

\begin{figure}[htp]
    \centering
    \includegraphics[width=1.0\linewidth]{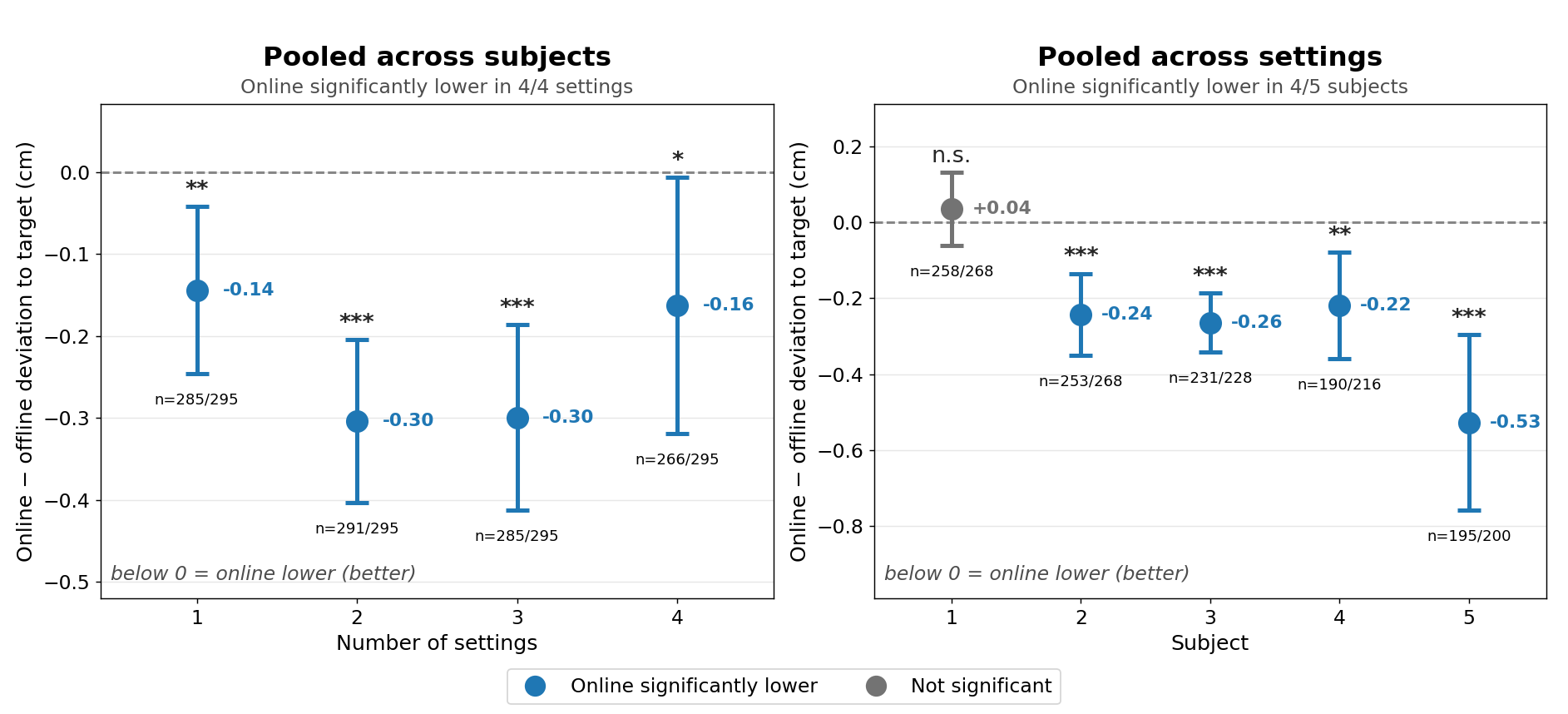}
    \caption{{\color{black}Online-offline gap in deviation to target for resolution reconfiguration. Each point is the linear mixed-effects estimate of the online effect (the $\beta_1$ coefficient, i.e. the
  mean online-offline difference in deviation to target) with its 95$\%$ confidence interval. Negative values indicate lower online deviation. Left: the model is fit separately at each resolution setting with subject as the grouping variable, so each estimate is the gap at that setting pooled across subjects. Right: the model is fit separately for each subject with setting as the grouping
  variable, so each estimate is that subject's
  gap pooled across settings.}}
    \label{fig:rr_offline_vs_online_analysis}
\end{figure}

\begin{figure}[htp]
    \centering
    \includegraphics[width=1.0\linewidth]{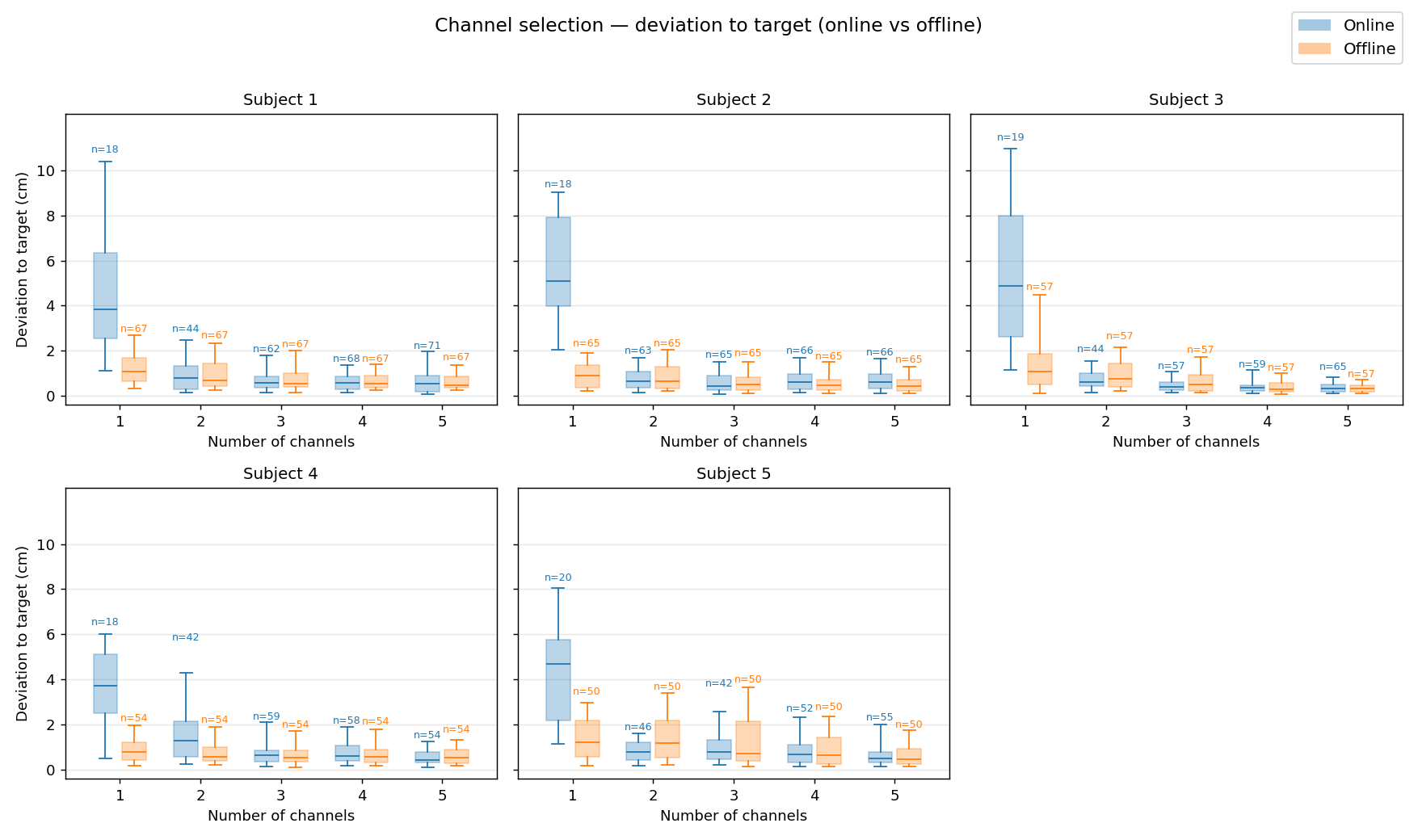}
    \caption{{\color{black}Online decoder vs. Offline decoder performance of five subjects given same channel selection settings. Note that in the 1-ch setting, offline decoders gives overly optimistic estimate.}}
    \label{fig:cs_offline_vs_online}
\end{figure}

\begin{figure}[htp]
    \centering
    \includegraphics[width=1.0\linewidth]{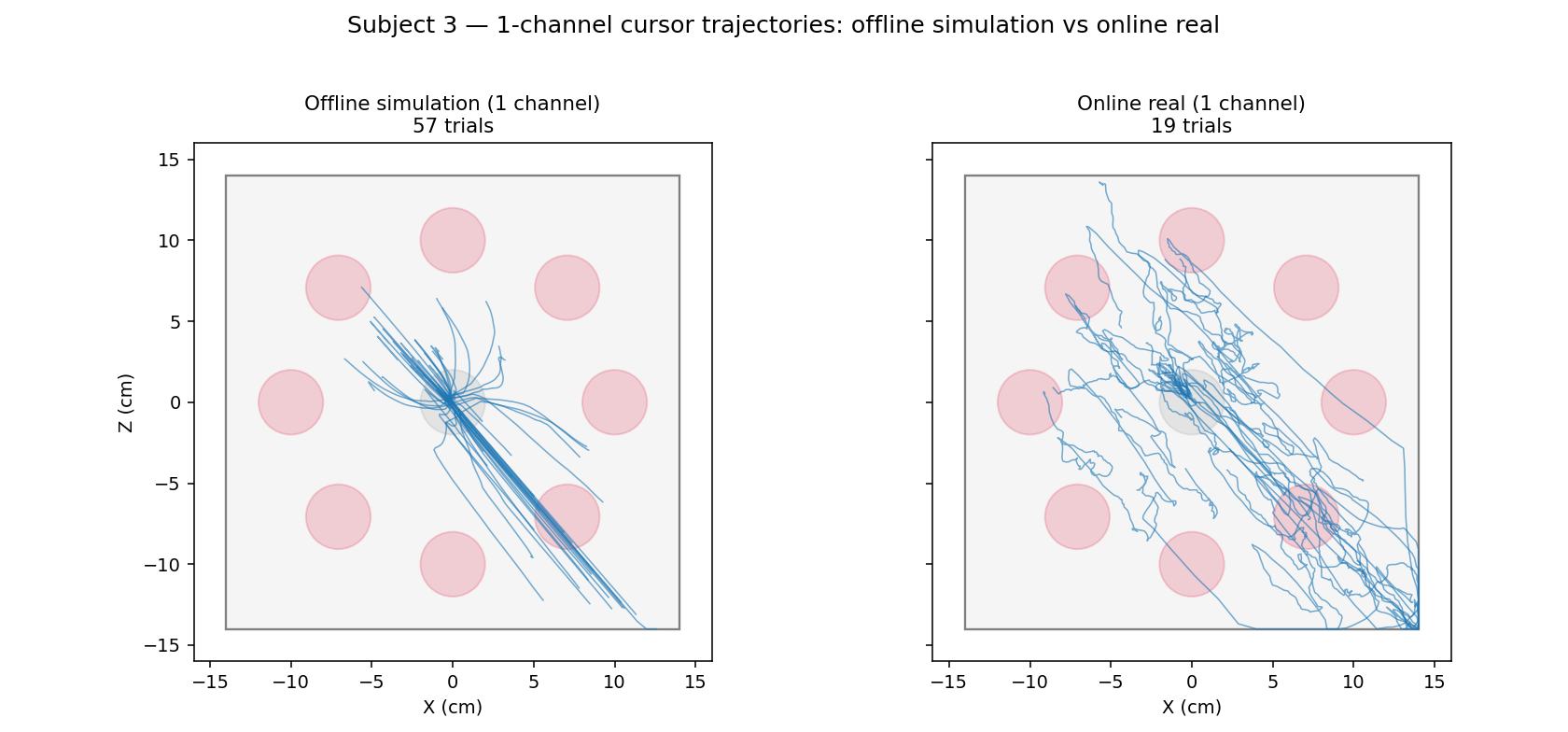}
    \caption{{\color{black}Example of cursor trajectories for 1-ch setting. The left one is the offline simulated trajectory and the right one is the online recorded trajectory. Both offline and online trajectories indicate that the cursor is moving on a 1-D manifold, while the online trajectory shows some local perturbations because the subject struggles to correct the cursor movement from cursor feedback.}}
    \label{fig:cursor_offline_vs_online}
\end{figure}

\begin{figure}[htp]
    \centering
    \includegraphics[width=1.0\linewidth]{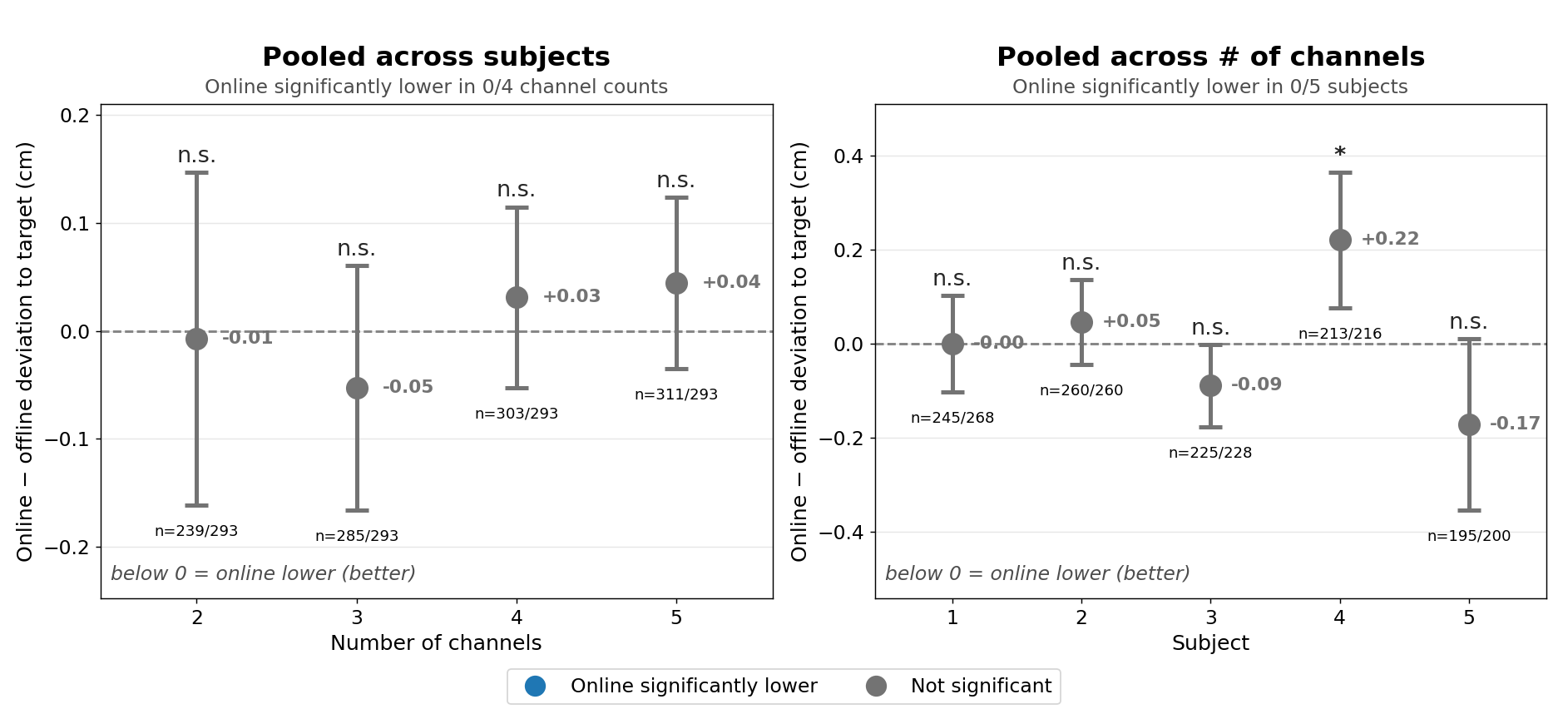}
    \caption{{\color{black}Online-offline gap in deviation to target for channel selection. Each point is the mixed-effects estimate of the online effect (online-offline difference in deviation to target) $\pm$ 95$\%$ CI, with negative values indicating lower
  (better) online deviation. Left: fit separately at each channel count with subject as the grouping
  variable, pooled across subjects. Right: fit separately for each subject with channel count as the
  grouping variable, pooled across channel counts. The single-channel configuration is excluded. The online decoder was not significantly lower at any channel count or in any subject. Subject 4 reached significance, but in the
  direction of worse online performance ($+0.22$ cm).}}
    \label{fig:cs_offline_vs_online_analysis}
\end{figure}

{\color{black}To quantify user adaptation to channel selection and resolution reconfiguration, we simulated offline decoder performance using the same configurations as the online experiment and compared the results. 

Offline performance was simulated by decoding EMG data from the 5-minute closed-loop baseline block (utilizing all five channels with 20 $\mu$Vrms added noise), incorporating channel selection and resolution reconfiguration. For channel selection, we removed the dropped channels from the baseline data and recalculated the steady-state Kalman filter matrices by omitting the corresponding rows in the observation and observation noise matrices. For resolution reconfiguration, we injected random Gaussian noise matching the per-channel noise floor of the online experiment and updated the observation noise matrix to solve the Kalman filter accordingly.

To evaluate decoder performance, we used "deviation to target", which is the same trajectory metric used to determine channel importance (see Methods: Online sEMG Cursor Control). This is defined as the average perpendicular distance between the simulated cursor trajectory and an ideal straight line from the start to the target. We deliberately chose this spatial metric over movement time because offline simulations rely on temporally fixed recorded data and cannot accurately capture dynamic, closed-loop changes in movement time.


Figure \ref{fig:rr_offline_vs_online} shows the decoder performance for online versus offline resolution reconfiguration for five subjects. The online median deviation to target was lower than offline across the majority of resolution settings. Linear mixed-effects models are used to quantify this performance gap while accounting for the non-independence of trials from the same subject and for the systematic differences of deviation to target across resolution settings. For trial $j$ belonging to group $i$, the deviation to target was modeled as:

\begin{equation}
  y_{ij} = \beta_0 + \beta_1\times\mathrm{Online}_{ij} + b_i + \epsilon_{ij},
  \quad b_i \sim \mathcal{N}(0,\sigma_b^{2}),\quad \epsilon_{ij}\sim\mathcal{N}(0,\sigma^{2}),
\end{equation}

where $\mathrm{Online}_{ij}=1$ for online trials and $0$ for offline trials, $\beta_{0}$ is a fixed intercept that represents the global average offline deviation to target, $b_i$ is a random intercept that adjusts the average performance of group $i$ (so repeated trials from the same group are not treated as independent), and $\epsilon_{ij}$ is residual noise. The fixed effect $\beta_1$ is the mean online-offline difference in deviation to target, with $\beta_1<0$ indicating better (lower) online performance. We fit this model two ways: separately at each resolution setting with subject as the grouping variable (random intercept per subject), and separately for each subject with setting as the grouping variable.
Figure \ref{fig:rr_offline_vs_online_analysis} demonstrates the estimated online-offline gap ($\beta_1$, with 95$\%$ confidence intervals). At every resolution setting, the online decoder achieved a significantly lower deviation to target than the offline decoder (all $p$-value $<0.05$). Per-subject analysis shows that, in 4 of the 5 subjects, the online decoder achieves a significantly lower deviation to target.

Figure \ref{fig:cs_offline_vs_online} compares decoder performance between online and offline channel selection. A notable performance gap exists in the single-channel (1-ch) configuration, where offline simulations significantly outperformed online results. This disparity arises because, in the online 1-ch condition, users lack dexterous 2-D control and frequently attempt corrective movements while constrained to a single axis. In contrast, the offline simulation predicts trajectories based on baseline recordings of successful trials, resulting in overly optimistic estimates. As illustrated in Figure \ref{fig:cursor_offline_vs_online}, while both trajectories move along a similar 1-D manifold, the online trajectory exhibits more perturbations as the subject struggles to adjust the cursor.
However, when restricting the analysis to configurations where users could successfully perform the online task ($>$1 channel, $>$90$\%$ target hits), the linear mixed-effects model described above indicated no significant difference between offline and online performance at any number of channels. In Fig. \ref{fig:cs_offline_vs_online_analysis}, per-subject analysis also shows only Subject 4 is significant but in the direction of worse online performance. Taken together, these findings suggest that resolution reconfiguration changes information content in a manner that is easier for users to adapt to than channel selection.}

\clearpage
\newpage
\section*{Non-Linear Classifier Channel Selection Results}

\begin{figure}[htp]
    \centering
    \includegraphics[width=\textwidth]{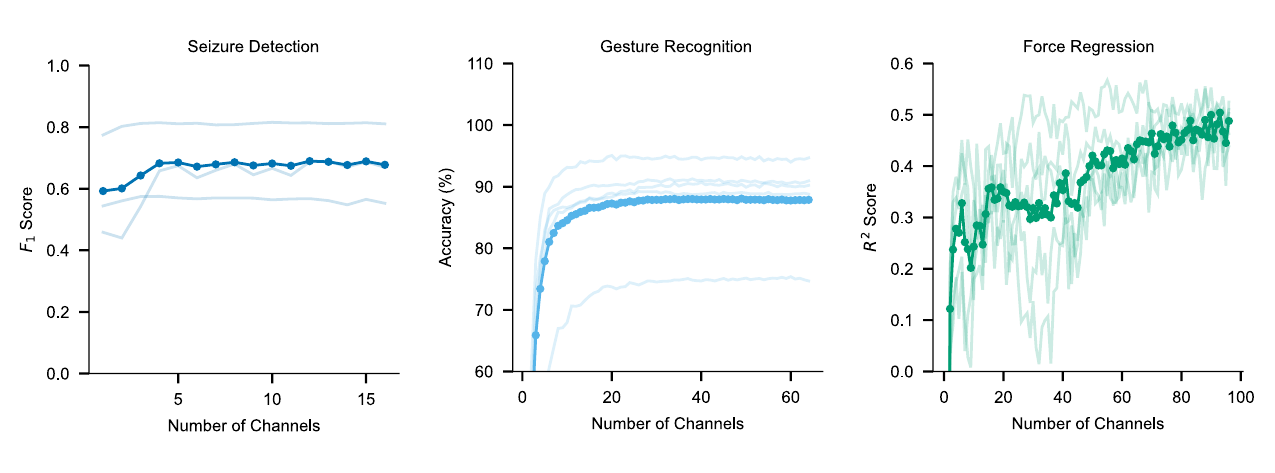}
    \caption{Channel selection results for nonlinear decoders using the same recursive channel selection strategy described in Results}
    \label{fig:nonlinear-channel-selection}
\end{figure}

\section*{Effect of Power Scaling Factor $<$4x}

\begin{figure}[htp]
    \centering
    \includegraphics[width=1.0\linewidth]{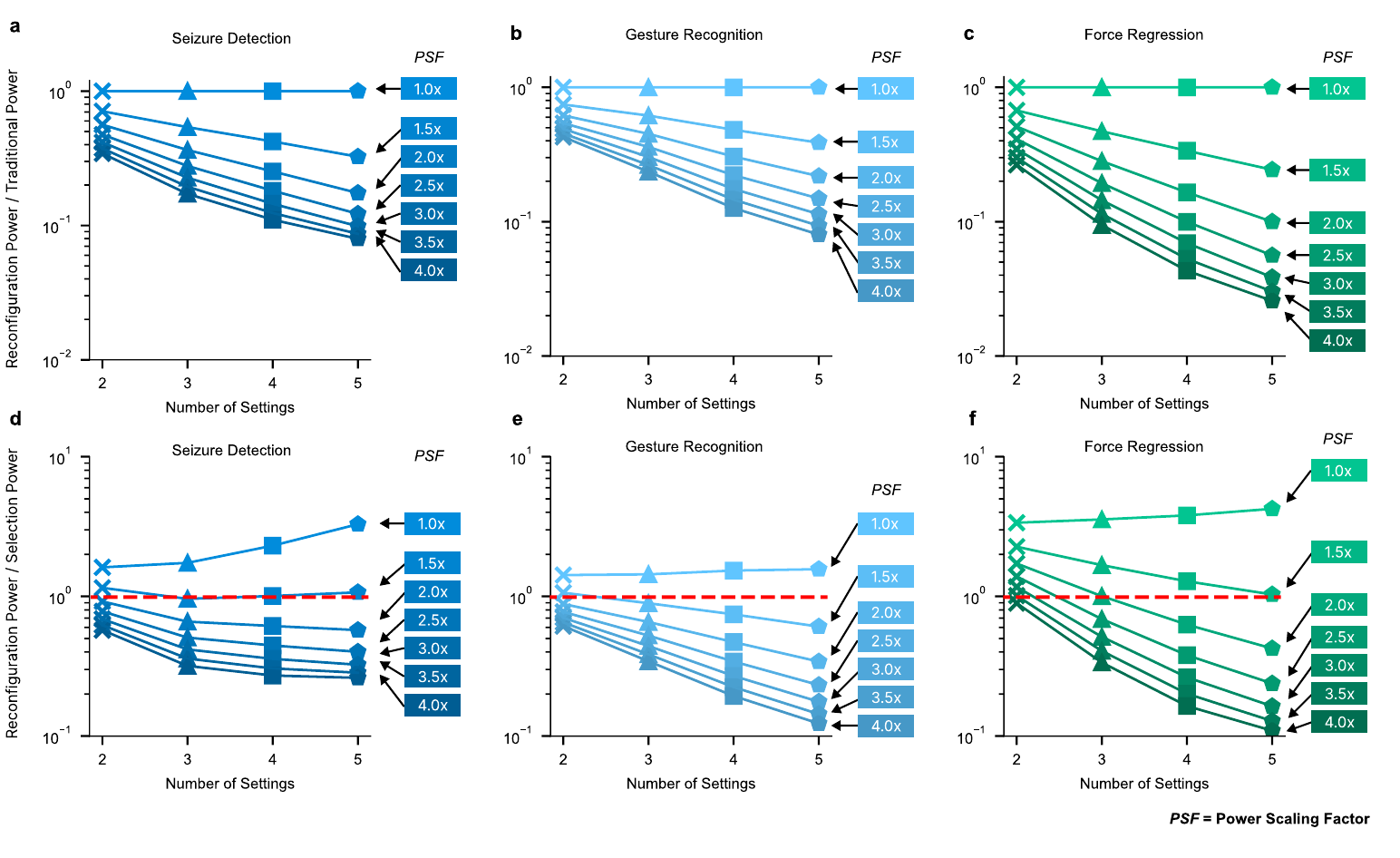}
    \caption{{\color{black}Subfigure (a)-(c) show the relative AFE power as a function of the number of settings, across different AFE power scaling factors. Subfigure (d)-(f) compare the channel reconfiguration power and sparse channel selection power at equivalent accuracies as a function of the number of settings across different AFE power scaling factors.}}
    \label{fig:scaling}
\end{figure}

In the Main body of this work, we assumed that halving channel resolution (i.e., reducing it by one digital bit) corresponds to a four-fold reduction in power consumption. In practice, however, biasing constraints, bandwidth requirements, and minimum transistor or passive component sizes can limit this scaling. Figure \ref{fig:scaling} compares power savings across different scaling factors: a value of 4x represents the ideal case used for the results in this paper, while a value of 1x indicates no power benefit from resolution reduction. Figure \ref{fig:scaling} a)-c) quantifies the power savings across scaling factors and resolution settings relative to a full array, and d)-e) compares the same settings relative to channel selection for equivalent accuracy. Our results show that for gesture recognition and force regression resolution reconfiguration will outperform channel selection, if the resolution scale factor is 2x or greater, and the number of settings is three or greater. 

Our analysis shows power scales by a factor of 4$\times$ for every bit of resolution, but even if power scaled by slightly less, we would still save power relative to channel selection. In a circuit implementation, practical design concerns will limit both the power scale factor and the number of unique settings the system can achieve. We compared performance across systems with two to five unique settings and compared the power savings across various scale factors. Our findings indicate that increasing the number of unique settings leads to diminishing returns and even three unique settings may save significant power when compared to channel selection for the gesture recognition and force regression tasks. At the same time, it is essential that the power scale factor be as high as possible. This indicates that designs should prioritize fewer settings with high power scaling factors, with a minimum three settings at a power scaling factor of at least 3$\times$.

\clearpage
\section*{Linear Classifier Hyperparameters}

The tables below contain the hyper parameter settings for the linear classifiers. For seizure detection, the linear classifier used was Logistic Regression, and the hyperparameters are listed in table \ref{tab:melbourne_lr_hyperparameters} and table \ref{tab:seizure_lr_hyperparameters}. For EMG Gesture recognition, the linear classifier used was also Logistic Regression, and the hyperparameters are listed in table \ref{tab:emg_lr_hyperparameters}. Finally, for the reach and grasp task, the linear decoder was a ridge regressor, for which the hyperparameters are outlined in table \ref{tab:ridge_hyperparameters}.

\begin{table}[h]
    \centering
    \begin{tabular}{l|r|r|r}
        \toprule
         & 1 & 2 & 3 \\
        $C$ &  &  &  \\
        \midrule
        0.010000 & 0.498 & 0.678 & 0.506 \\
        0.100000 & 0.521 & 0.730 & 0.512 \\
        1.000000 & 0.529 & 0.734 & 0.512 \\
        10.000000 & 0.531 & 0.734 & 0.512 \\
        100.000000 & 0.531 & 0.734 & 0.512 \\
        \bottomrule
    \end{tabular}
    \caption{{\color{black}Logistic regression hyperparameter tuning for the seizure detection task using the Melbourne dataset. $C$ represents the L2 penalty for the logistic regression classifier.}}
    \label{tab:melbourne_lr_hyperparameters}
\end{table}

\begin{table}[h]
    \centering
    \begin{tabular}{l|r|r|r|r|r|r|r|r}
\hline
$C$ & 1 & 2 & 3 & 4 & 5 & 6 & 7 & 8 \\
\hline
0.01 & 0.997 & 0.965 & 0.990 & 0.630 & 0.995 & 0.874 & 0.935 & 0.906 \\
0.10 & 0.999 & 0.984 & 0.999 & 0.972 & 1.000 & 0.994 & 0.985 & 0.968 \\
1.00 & 0.998 & 0.991 & 0.998 & 0.979 & 0.999 & 0.998 & 0.994 & 0.985 \\
10.0 & 0.998 & 0.990 & 0.996 & 0.997 & 0.996 & 0.995 & 1.000 & 0.982 \\
100 & 0.998 & 0.991 & 0.991 & 0.996 & 0.994 & 0.994 & 0.999 & 0.973 \\
\end{tabular}

\begin{tabular}{l|r|r|r|r|r|r|r|r}
\hline
$C$ & 9 & 10 & 11 & 12 & 13 & 14 & 15 & 16 \\
\hline
0.01 & 0.992 & 0.988 & 0.981 & 0.940 & 0.867 & 0.954 & 0.942 & 0.699 \\
0.10 & 0.992 & 0.995 & 1.000 & 0.970 & 0.968 & 0.986 & 0.980 & 0.970 \\
1.00 & 0.999 & 0.994 & 1.000 & 0.975 & 0.973 & 0.987 & 0.986 & 0.934 \\
10.0 & 0.998 & 0.992 & 1.000 & 0.971 & 0.966 & 0.969 & 0.984 & 0.914 \\
100 & 0.998 & 0.992 & 1.000 & 0.972 & 0.959 & 0.949 & 0.984 & 0.862 \\
\bottomrule
\end{tabular}

\begin{tabular}{l|r|r|r|r|r|r|r|r}
\toprule
$C$ & 17 & 18 & 19 & 20 & 21 & 22 & 23 & 24 \\
\midrule
0.01 & 0.711 & 0.911 & 0.952 & 0.947 & 0.793 & 0.984 & 0.984 & 0.863 \\
0.10 & 0.950 & 0.944 & 0.980 & 0.997 & 0.999 & 0.996 & 0.994 & 0.901 \\
1.00 & 0.981 & 0.974 & 0.997 & 0.998 & 0.997 & 0.997 & 0.994 & 0.894 \\
10.0 & 0.965 & 0.978 & 0.973 & 0.996 & 0.987 & 0.999 & 0.983 & 0.892 \\
100 & 0.974 & 0.978 & 0.987 & 0.992 & 0.973 & 0.994 & 0.980 & 0.886 \\
\bottomrule
\end{tabular}
    
    \caption{{\color{black}Logistic regression hyperparameter tuning for the seizure detection task in CHB-MIT. $C$ represents the L2 penalty for the logistic regression classifier.}}
    \label{tab:seizure_lr_hyperparameters}
\end{table}

\begin{table}[h]
    \centering
    \begin{tabular}{c|c|c|c|c|c}
    \toprule
        $C$ & Patient 1 & Patient 2 & Patient 3 & Patient 4 & Patient 5 \\
    \midrule
        $0.01$ & 0.882 & 0.912 & 0.869 & 0.834 & 0.668\\
        $0.1$ & 0.930 & 0.947 & 0.904 & 0.899 & 0.781\\
        $1$ & 0.962 & 0.964 & 0.921 & 0.931 & 0.858\\
        $10$ & 0.968 & 0.965 & 0.920 & 0.944 & 0.881\\
        $100$ & 0.965 & 0.962 & 0.918 & 0.943 & 0.877\\
    \bottomrule
    \end{tabular}
    \caption{Logistic regression hyperparameter tuning for the gesture classification task. $C$ represents the L2 penalty for the logistic regression classifier.}
    \label{tab:emg_lr_hyperparameters}
\end{table}

\begin{table}[h]
    \centering
    \begin{tabular}{c|c}
    \toprule
        $\alpha$ & $R^2$ CV Score \\
    \midrule
        $1.00 \cdot 10^{-5}$ & 0.365 \\
        $1.29 \cdot 10^{-4}$ & 0.365 \\
        $1.67 \cdot 10^{-3}$ & 0.365 \\
        $2.15 \cdot 10^{-2}$ & 0.365 \\
        $2.78 \cdot 10^{-1}$ & 0.365 \\
        $3.59 \cdot 10^{0}$  & 0.365 \\
       $ 4.64 \cdot 10^{1}$  & 0.365 \\ 
        $5.99 \cdot 10^{2}$  & 0.365 \\ 
        $7.74 \cdot 10^{3}$  & 0.360 \\ 
        $1.00 \cdot 10^{5}$  & 0.294 \\
    \bottomrule
    \end{tabular}
    \caption{Ridge hyperparameter tuning for the reach-to-grasp task. The cross-validated $R^2$ score is fairly constant across values of $\alpha$. $\alpha$ represents the L2 penalty for the Ridge regressor}
    \label{tab:ridge_hyperparameters}
\end{table}


\clearpage
\newpage
\section*{Detailed Results on Weights and Channel Assignments}

The plots below show detailed results on the distribution of importance scores across cross validations and their corresponding channel assignments for the case with 5 different settings for each ADC. 

\begin{figure}[h]
    \centering
    \includegraphics[width=\textwidth]{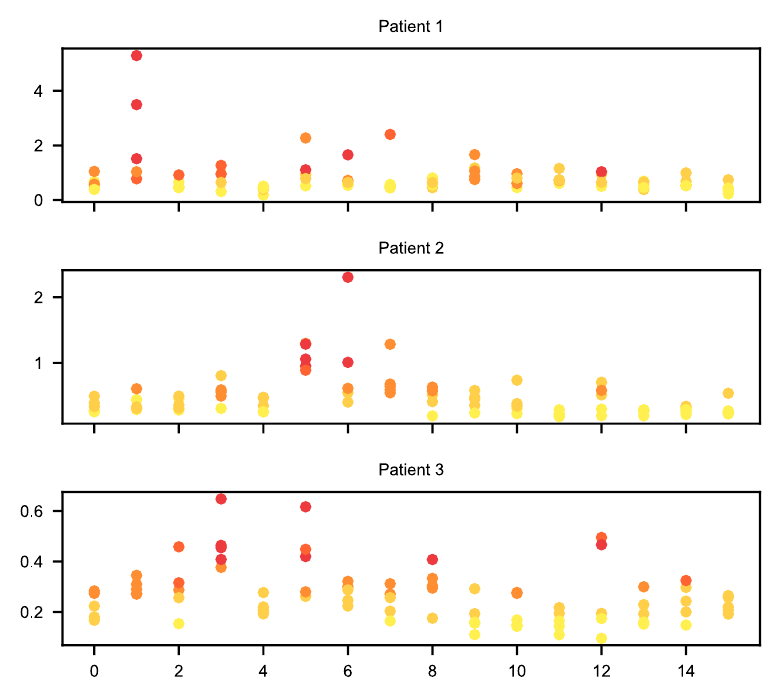}
    \caption{{\color{black}The distribution of importance scores and the corresponding channel settings for every patient in the Melbourne dataset for a linear classifier. Each dot corresponds to an importance score for a cross validation.}}
\end{figure}

\begin{figure}[h]
    \centering
    \includegraphics[width=\textwidth]{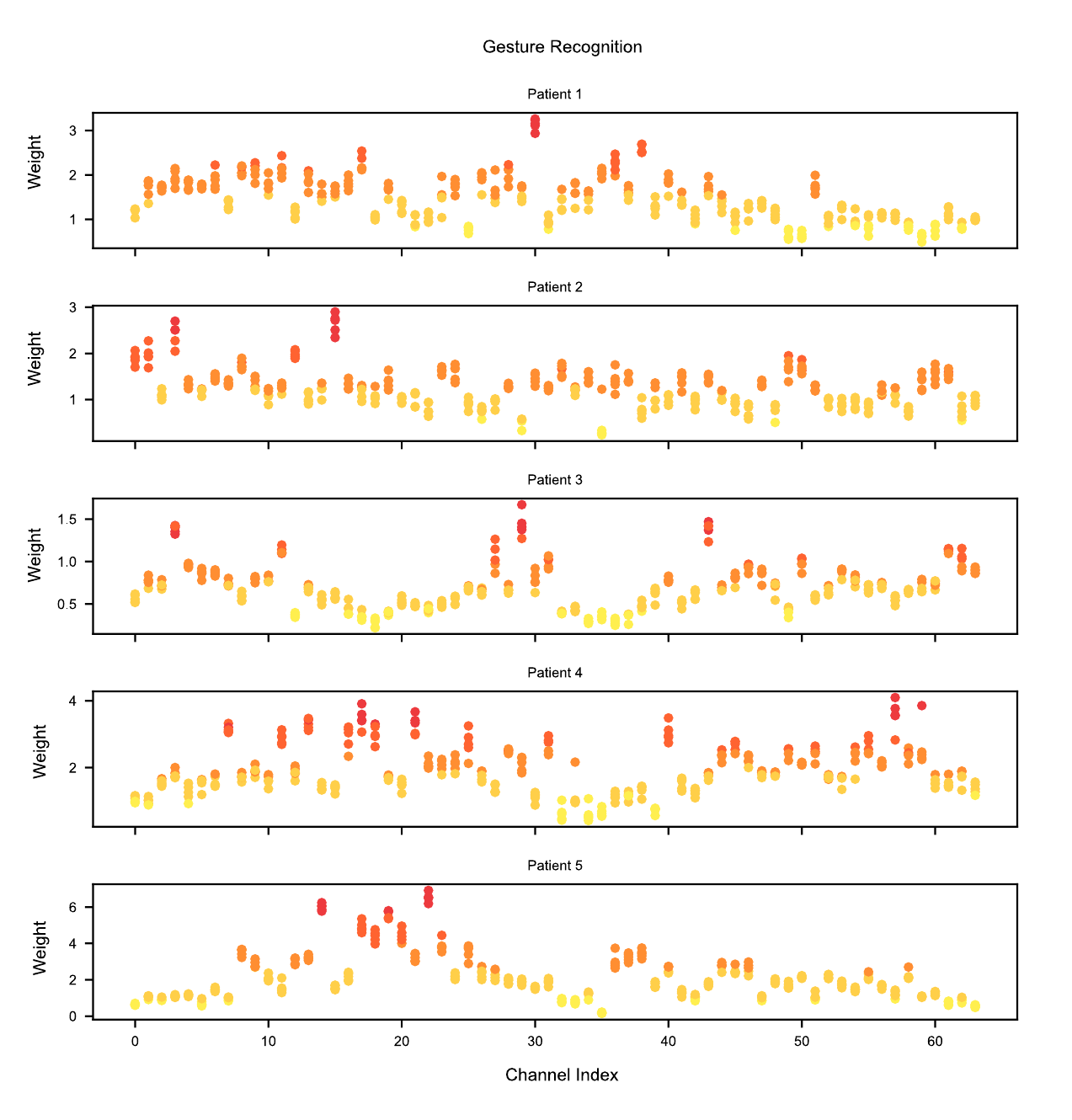}
    \caption{{\color{black}The distribution of importance scores and the corresponding channel settings for every patient in the EMG gesture recognition dataset for a linear classifier. Each dot corresponds to an importance score for a cross validation.}}
\end{figure}

\begin{figure}[h]
    \centering
    \includegraphics[width=\textwidth]{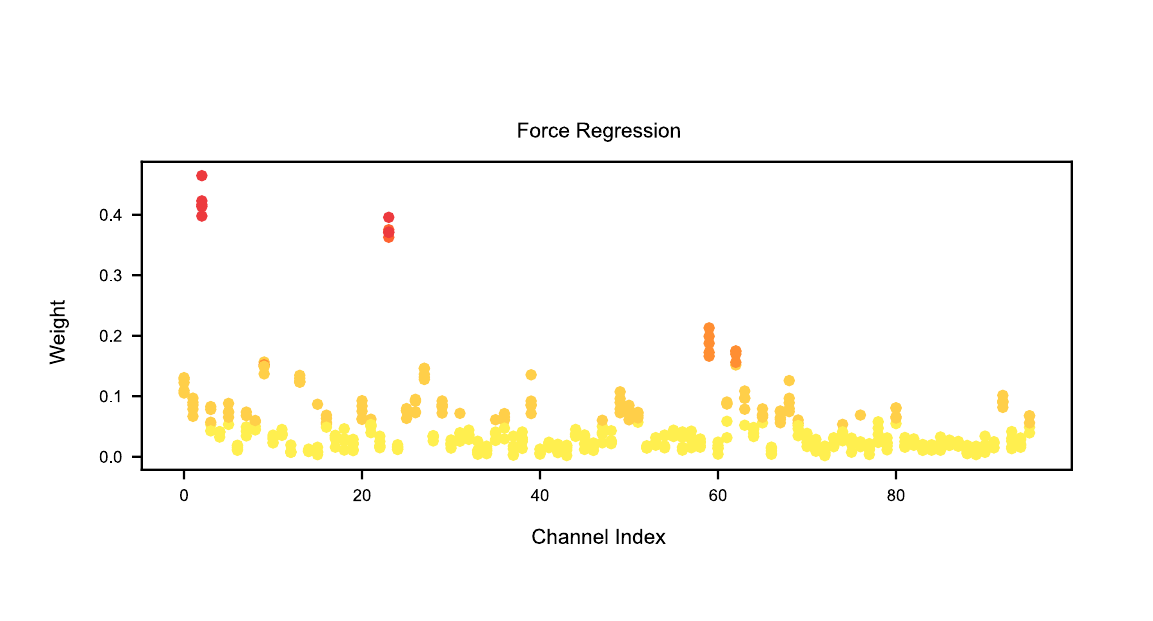}
    \caption{{\color{black}The distribution of importance scores and the corresponding channel settings for every cross validation in the reach and grasp dataset for a linear classifier.}}
\end{figure}

\clearpage
\newpage
\section*{Linear Classifier Settings Monte-Carlo}

\begin{figure}[htp]
    \centering
    \includegraphics[width=0.8\textwidth]{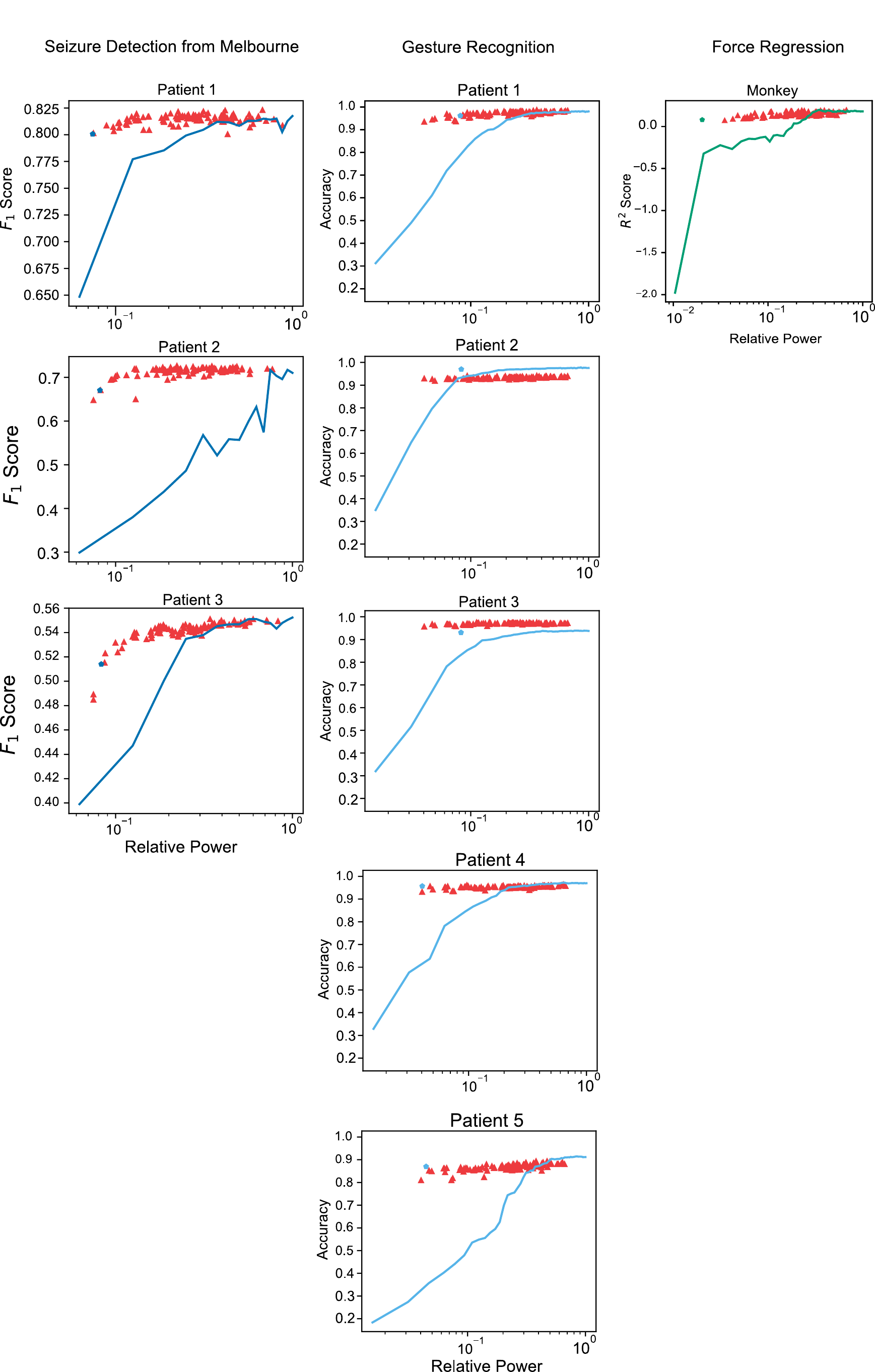}
    \caption{{\color{black}Monte Carlo results for each task. For the Monte Carlo settings assignment, channels are first ranked by importance. Then, four sorted random integers from range $[$1, \# of channels$]$ are chosen to be the bin edges. This randomizes the bin sizes and assignments while preserving the relative importance of each channel. Results are shown only for a single patient with five settings enabled.}}
    \label{fig:monte_carlo}
\end{figure}

{\color{black}To evaluate the effectiveness of our importance-to-settings mapping, we apply a Monte Carlo approach: channel importances are extracted and ranked once, then randomly sampled bin edges with replacement are used to assign channels to settings, simulating the full space of valid mapping strategies. All sampled mappings enforce two invariants: (1) the highest and lowest importance channels always map to the highest and lowest settings respectively, and (2) importance order is strictly preserved, so no higher-importance channel receives a lower setting than a less important one.}

This Monte Carlo sweep reveals where our linear mapping sits relative to all valid strategies. As shown in the main body, our mapping achieves strong power efficiency, saving an order of magnitude over full arrays and 3x to 5x over channel selection, and consistently lands on the power-efficiency frontier. Critically, our linear mapping achieves this without random search: the mapping is derived directly from importance weights. While better strategies may exist outside of this constrained method (as noted in the Discussion), the results here confirm that enforcing importance-to-settings monotonicity is sufficient to achieve strong empirical power efficiency.

\newpage
\section*{Non-linear Classifier Hyperparameters}

\begin{table}[h]
    \centering
    \begin{tabular}{lrlrrr}
    \toprule
    criterion & n\_estimators & max\_features & 1 & 2 & 3 \\
    \midrule
    entropy & 64 & 1.000 & 0.737 & 0.584 & 0.559 \\
    entropy & 64 & sqrt & 0.720 & 0.612 & 0.544 \\
    entropy & 128 & 1.000 & 0.735 & 0.582 & 0.560 \\
    entropy & 128 & sqrt & 0.720 & 0.626 & 0.547 \\
    entropy & 256 & 1.000 & 0.735 & 0.572 & 0.562 \\
    entropy & 256 & sqrt & 0.720 & 0.622 & 0.548 \\
    gini & 64 & 1.000 & 0.717 & 0.577 & 0.552 \\
    gini & 64 & sqrt & 0.716 & 0.587 & 0.533 \\
    gini & 128 & 1.000 & 0.718 & 0.573 & 0.553 \\
    gini & 128 & sqrt & 0.714 & 0.598 & 0.536 \\
    gini & 256 & 1.000 & 0.719 & 0.577 & 0.553 \\
    gini & 256 & sqrt & 0.717 & 0.600 & 0.537 \\
    \bottomrule
    \end{tabular}
    \caption{{\color{black}Nonlinear hyparameter sweep for Melbourne using a Random Forest classifier.}}
    \label{tab:melbourne_rf_hyperparams}
\end{table}

\begin{table}[hp]
    \centering
    \begin{tabular}{l|r|l|r|r|r|r|r|r|r|r}
\toprule
\hline
criterion & n\_estimators & max\_features & 1 & 2 & 3 & 4 & 5 & 6 & 7 & 8 \\
\hline
\midrule
entropy & 64 & 1.000 & 0.983 & 0.986 & 0.990 & 0.972 & 0.991 & 0.862 & 0.979 & 0.945 \\
entropy & 64 & sqrt & 0.982 & 0.976 & 0.991 & 0.965 & 0.993 & 0.810 & 0.977 & 0.928 \\
entropy & 128 & 1.000 & 0.984 & 0.988 & 0.991 & 0.973 & 0.993 & 0.867 & 0.981 & 0.946 \\
entropy & 128 & sqrt & 0.982 & 0.977 & 0.988 & 0.969 & 0.993 & 0.823 & 0.977 & 0.934 \\
entropy & 256 & 1.000 & 0.984 & 0.995 & 0.993 & 0.968 & 0.992 & 0.861 & 0.980 & 0.945 \\
entropy & 256 & sqrt & 0.982 & 0.977 & 0.990 & 0.968 & 0.992 & 0.831 & 0.979 & 0.929 \\
gini & 64 & 1.000 & 0.983 & 0.995 & 0.991 & 0.967 & 0.992 & 0.861 & 0.977 & 0.930 \\
gini & 64 & sqrt & 0.982 & 0.965 & 0.987 & 0.957 & 0.987 & 0.816 & 0.973 & 0.917 \\
gini & 128 & 1.000 & 0.982 & 0.990 & 0.990 & 0.972 & 0.992 & 0.864 & 0.980 & 0.932 \\
gini & 128 & sqrt & 0.981 & 0.978 & 0.987 & 0.971 & 0.993 & 0.798 & 0.977 & 0.919 \\
gini & 256 & 1.000 & 0.983 & 0.990 & 0.990 & 0.971 & 0.992 & 0.867 & 0.981 & 0.932 \\
gini & 256 & sqrt & 0.981 & 0.978 & 0.988 & 0.964 & 0.992 & 0.809 & 0.977 & 0.921 \\
\bottomrule
\end{tabular}

\begin{tabular}{l|r|l|r|r|r|r|r|r|r|r}
\toprule
\hline
criterion & n\_estimators & max\_features & 9 & 10 & 11 & 12 & 13 & 14 & 15 & 16 \\
\hline
\midrule
entropy & 64 & 1.000 & 0.999 & 0.993 & 0.987 & 0.822 & 0.861 & 0.844 & 0.968 & 0.318 \\
entropy & 64 & sqrt & 1.000 & 0.987 & 0.984 & 0.776 & 0.810 & 0.797 & 0.944 & 0.215 \\
entropy & 128 & 1.000 & 0.999 & 0.993 & 0.984 & 0.819 & 0.859 & 0.841 & 0.969 & 0.290 \\
entropy & 128 & sqrt & 1.000 & 0.987 & 0.978 & 0.774 & 0.820 & 0.792 & 0.943 & 0.179 \\
entropy & 256 & 1.000 & 0.999 & 0.993 & 0.984 & 0.819 & 0.861 & 0.840 & 0.969 & 0.312 \\
entropy & 256 & sqrt & 1.000 & 0.988 & 0.981 & 0.772 & 0.822 & 0.793 & 0.945 & 0.201 \\
gini & 64 & 1.000 & 1.000 & 0.993 & 0.990 & 0.794 & 0.820 & 0.834 & 0.961 & 0.318 \\
gini & 64 & sqrt & 0.987 & 0.989 & 0.991 & 0.773 & 0.752 & 0.719 & 0.945 & 0.076 \\
gini & 128 & 1.000 & 1.000 & 0.993 & 0.975 & 0.800 & 0.836 & 0.836 & 0.963 & 0.310 \\
gini & 128 & sqrt & 1.000 & 0.989 & 0.981 & 0.765 & 0.804 & 0.774 & 0.938 & 0.168 \\
gini & 256 & 1.000 & 0.999 & 0.993 & 0.981 & 0.797 & 0.832 & 0.839 & 0.963 & 0.336 \\
gini & 256 & sqrt & 1.000 & 0.988 & 0.981 & 0.764 & 0.799 & 0.771 & 0.940 & 0.151 \\
\bottomrule
\end{tabular}

\begin{tabular}{l|r|l|r|r|r|r|r|r|r|r}
\toprule
\hline
criterion & n\_estimators & max\_features & 17 & 18 & 19 & 20 & 21 & 22 & 23 & 24 \\
\hline
\midrule
entropy & 64 & 1.000 & 0.948 & 0.911 & 0.974 & 0.936 & 0.930 & 0.991 & 0.973 & 0.553 \\
entropy & 64 & sqrt & 0.907 & 0.892 & 0.959 & 0.899 & 0.911 & 0.983 & 0.972 & 0.536 \\
entropy & 128 & 1.000 & 0.937 & 0.916 & 0.979 & 0.928 & 0.924 & 0.988 & 0.975 & 0.555 \\
entropy & 128 & sqrt & 0.924 & 0.887 & 0.958 & 0.904 & 0.903 & 0.984 & 0.973 & 0.539 \\
entropy & 256 & 1.000 & 0.946 & 0.920 & 0.974 & 0.935 & 0.927 & 0.991 & 0.975 & 0.564 \\
entropy & 256 & sqrt & 0.927 & 0.881 & 0.952 & 0.902 & 0.912 & 0.983 & 0.975 & 0.543 \\
gini & 64 & 1.000 & 0.927 & 0.912 & 0.973 & 0.880 & 0.917 & 0.988 & 0.974 & 0.531 \\
gini & 64 & sqrt & 0.938 & 0.889 & 0.958 & 0.875 & 0.886 & 0.976 & 0.976 & 0.563 \\
gini & 128 & 1.000 & 0.939 & 0.906 & 0.977 & 0.882 & 0.918 & 0.983 & 0.974 & 0.524 \\
gini & 128 & sqrt & 0.914 & 0.882 & 0.952 & 0.860 & 0.903 & 0.982 & 0.973 & 0.520 \\
gini & 256 & 1.000 & 0.940 & 0.912 & 0.967 & 0.881 & 0.921 & 0.990 & 0.974 & 0.532 \\
gini & 256 & sqrt & 0.910 & 0.882 & 0.958 & 0.863 & 0.909 & 0.982 & 0.974 & 0.524 \\
\bottomrule
\end{tabular}
    
    \caption{{\color{black}Random Forest hyperparameter tuning for the seizure detection task from CHB-MIT.}}
    \label{tab:seizure_rf_hyperparameters}
\end{table}

\begin{table}[h]
    \centering
    \begin{tabular}{c|c|c|c|c|c}
    \toprule
        Number of Estimators & Patient 1 & Patient 2 & Patient 3 & Patient 4 & Patient 5 \\
    \midrule
        $10$ & 0.831 & 0.925 & 0.806 & 0.830 & 0.615\\
        $100$ & 0.837 & 0.928 & 0.806 & 0.863 & 0.642\\
        $500$ & 0.837 & 0.923 & 0.811 & 0.864 & 0.651\\
    \bottomrule
    \end{tabular}
    \caption{Random forest hyperparameter tuning for the gesture classification task. The number of estimators parameter represents the number of trees used in the random forest classifier.}
    \label{tab:emg_rf_hyperparameters}
\end{table}

\begin{table}[h]
    \centering
    \begin{tabular}{c | c|c}
    \toprule
        Layer 1 Size & Layer 2 Size & $R^2$ CV Score \\
    \midrule
48 & 48 & 0.438 \\
48 & 96 & 0.397 \\
48 & 192 & 0.387 \\
48 & 384 & 0.388 \\
48 & 768 & 0.371 \\
96 & 48 & 0.413 \\
96 & 96 & 0.386 \\
96 & 192 & 0.405 \\
96 & 384 & 0.406 \\
96 & 768 & 0.419 \\
192 & 48 & 0.45 \\
192 & 96 & 0.437 \\
192 & 192 & 0.438 \\
192 & 384 & 0.457 \\
192 & 768 & 0.422 \\
384 & 48 & 0.437 \\
384 & 96 & 0.412 \\
384 & 192 & 0.427 \\
384 & 384 & 0.442 \\
384 & 768 & 0.452 \\
768 & 48 & 0.443 \\
768 & 96 & 0.445 \\
768 & 192 & 0.426 \\
768 & 384 & 0.428 \\
768 & 768 & 0.442 \\
\bottomrule
    \end{tabular}
    \caption{Neural Network hyperparameter tuning for the reach-to-grasp task. The Neural Network consists of two fully connected layers, each followed by a BatchNorm and ReLU for the activation. The layer sizes are the tuned hyperparameters. Generally, the scores were better for larger values of the inner layers. However, since all scores were reasonably similar, and for training time and memory constraints, we choose a moderate performance of $(192, 96)$. }
    \label{tab:nn_hyperparameters}
\end{table}

\clearpage
\newpage
\section*{Full Results by Patient}

In this section, we provide the power savings relative to a traditional recording array where every channel is always active at a fixed noise floor for both sparse channel selection and resolution adjustment. The savings are provided on a per-patient basis and averaged across the cross validations for the seizure detection and gesture recognition datasets. Table \ref{tab:seizure}, table \ref{tab:gesture}, and table \ref{tab:grasp} provide the results for seizure detection, gesture recognition, and the reach and grasp task with a linear classifier respectively. Table \ref{tab:seizure_nl}, table \ref{tab:gesture_nl} and table \ref{tab:grasp_nl} provide the same results with non-linear classifiers. We note that for gesture recognition and for the reach and grasp task, resolution adjustment always provides power savings over channel selection (indicated by a positive ratio). However, for seizure detection, resolution adjustment only provides a power advantage over channel selection for some patients.

\begin{table}[h]
    \centering
    \begin{tabular}{lrrrr}
\toprule
 & CS & RR & Ratio & Savings at Eq. Perf. \\
\midrule
0 & 8.000 & 13.474 & 1.684 & 4.211 \\
1 & 1.333 & 12.263 & 9.200 & 9.198 \\
2 & 4.000 & 12.047 & 3.012 & 3.012 \\
\bottomrule
\end{tabular}

    \caption{{\color{black}For seizure detection from the Melbourne dataset, we show the Channel Selection power saving for $\Delta = 0.05$, the Resolution Reconfiguration power for $\Delta = 0.05$, the ratio between these two, and the power savings of Resolution Reconfiguration at $\Delta = 0.05$ divided by the power savings of CS \textit{at equivalent performance}  for five settings.}}
    \label{tab:melbourne}
\end{table}

\begin{table}[h]
    \centering
    \begin{tabular}{lrrrr}
\toprule
 & CS & RR & Ratio & Savings at Eq. Perf. \\
\midrule
0 & 1.692 & 13.020 & 7.695 & 10.061 \\
1 & 22.000 & 8.057 & 0.366 & 0.732 \\
2 & 2.000 & 8.771 & 4.386 & 4.385 \\
3 & 1.000 & 1.000 & 1.000 & 1.000 \\
4 & 4.400 & 6.945 & 1.578 & 6.945 \\
5 & 1.048 & 1.000 & 0.954 & 1.000 \\
6 & 22.000 & 6.970 & 0.317 & 3.168 \\
7 & 3.143 & 9.306 & 2.961 & 3.384 \\
8 & 5.500 & 7.239 & 1.316 & 1.316 \\
9 & 22.000 & 6.912 & 0.314 & 0.314 \\
10 & 22.000 & 9.434 & 0.429 & 1.286 \\
11 & 2.000 & 3.293 & 1.646 & 1.796 \\
12 & 2.250 & 6.606 & 2.936 & 4.771 \\
13 & 3.143 & 5.075 & 1.615 & 1.615 \\
14 & 8.667 & 7.051 & 0.814 & 1.356 \\
15 & 1.385 & 3.918 & 2.829 & 3.228 \\
16 & 4.500 & 3.556 & 0.790 & 0.790 \\
17 & 6.000 & 6.169 & 1.028 & 1.028 \\
18 & 3.000 & 4.085 & 1.362 & 1.816 \\
19 & 4.400 & 12.656 & 2.876 & 5.178 \\
20 & 2.000 & 4.403 & 2.202 & 4.310 \\
21 & 5.500 & 5.379 & 0.978 & 1.712 \\
22 & 2.750 & 9.846 & 3.580 & 4.923 \\
23 & 5.500 & 7.626 & 1.387 & 2.773 \\
\bottomrule
\end{tabular}

    \caption{{\color{black}For seizure detection in the CHB-MIT using a logistic regression, computed savings over a traditional array for channel selection, resolution readjustment, and the ratio between resolution readjustment and channel selection, and the power comparison at equivalent performance for five settings.}}
    \label{tab:seizure}
\end{table}

\begin{table}[h]
    \centering
    \begin{tabular}{lrrrr}
\toprule
 & CS & RR & Ratio & Savings at Eq. Perf. \\
\midrule
0 & 3.556 & 16.374 & 4.605 & 11.513 \\
1 & 8.000 & 13.388 & 1.674 & 5.439 \\
2 & 4.923 & 12.465 & 2.532 & 9.544 \\
3 & 3.048 & 8.773 & 2.878 & 6.854 \\
4 & 1.641 & 14.129 & 8.610 & 10.376 \\
\bottomrule
\end{tabular}

    \caption{For gesture recognition using a logistic regression, computed savings over a traditional array for channel selection, resolution readjustment, and the ratio between resolution readjustment and channel selection.}
    \label{tab:gesture}
\end{table}

\begin{table}[h]
    \centering
    \begin{tabular}{lrrrr}
\toprule
 & CS & RR & Ratio & Savings at Eq. Perf. \\
\midrule
0 & 3.692 & 10.894 & 2.951 & 3.064 \\
1 & 3.556 & 22.261 & 6.260 & 6.725 \\
2 & 5.333 & 34.420 & 6.454 & 6.454 \\
3 & 4.174 & 22.505 & 5.392 & 5.861 \\
4 & 5.647 & 34.276 & 6.070 & 7.141 \\
\bottomrule
\end{tabular}

    \caption{For force regression using a ridge regression, computed savings over a traditional array for channel selection, resolution readjustment, and the ratio between resolution readjustment and channel selection.}
    \label{tab:grasp}
\end{table}

\begin{table}[]
    \centering
    \begin{tabular}{lrrrr}
\toprule
& CS & RR & Ratio & Savings at Eq. Perf. \\
\midrule
0 & 16.000 & 14.629 & 0.914 & 1.829 \\
1 & 4.000 & 9.460 & 2.365 & 2.365 \\
2 & 16.000 & 15.114 & 0.945 & 2.834 \\
\bottomrule
\end{tabular}

    \caption{{\color{black}Per patient results for seizure detection from the Melbourne dataset using Random Forest classifier.}}
    \label{tab:melbourne_nl}
\end{table}

\begin{table}[h]
    \centering
    \begin{tabular}{lrrrr}
\toprule
& CS & RR & Ratio & Savings at Eq. Perf. \\
\midrule
0 & 5.500 & 19.373 & 3.522 & 19.373 \\
1 & 22.000 & 17.655 & 0.803 & 2.408 \\
2 & 7.333 & 16.776 & 2.288 & 3.050 \\
3 & 22.000 & 18.227 & 0.828 & 1.657 \\
4 & 22.000 & 15.946 & 0.725 & 1.450 \\
5 & 7.333 & 3.520 & 0.480 & 0.480 \\
6 & 5.500 & 11.022 & 2.004 & 9.891 \\
7 & 7.333 & 15.732 & 2.145 & 5.006 \\
8 & 22.000 & 7.944 & 0.361 & 0.361 \\
9 & 22.000 & 14.404 & 0.655 & 0.655 \\
10 & 11.000 & 7.652 & 0.696 & 0.696 \\
11 & 22.000 & 19.986 & 0.908 & 0.908 \\
12 & 9.000 & 16.406 & 1.823 & 1.823 \\
13 & 22.000 & 2.919 & 0.133 & 0.133 \\
14 & 26.000 & 23.687 & 0.911 & 1.822 \\
15 & 18.000 & 7.481 & 0.416 & 0.416 \\
16 & 1.385 & 6.545 & 4.726 & 6.545 \\
17 & 18.000 & 15.890 & 0.883 & 0.883 \\
18 & 6.000 & 15.945 & 2.658 & 3.543 \\
19 & 11.000 & 16.837 & 1.531 & 2.296 \\
20 & 11.000 & 10.957 & 0.996 & 0.996 \\
21 & 22.000 & 17.171 & 0.780 & 0.780 \\
22 & 7.333 & 14.404 & 1.964 & 2.619 \\
23 & 22.000 & 20.277 & 0.922 & 2.765 \\
\bottomrule
\end{tabular}

    \caption{{\color{black}For seizure detection from CHB-MIT using a random forest, computed savings over a traditional array for channel selection, resolution readjustment, and the ratio between resolution readjustment and channel selection.}}
    \label{tab:seizure_nl}
\end{table}

\begin{table}[h]
    \centering
    \begin{tabular}{lrrrr}
\toprule
 & CS & RR & Ratio & Savings at Eq. Perf. \\
\midrule
0 & 8.000 & 9.388 & 1.174 & 4.401 \\
1 & 10.667 & 12.127 & 1.137 & 3.600 \\
2 & 10.667 & 22.315 & 2.092 & 7.671 \\
3 & 10.667 & 28.336 & 2.656 & 11.512 \\
4 & 5.818 & 10.448 & 1.796 & 5.550 \\
\bottomrule
\end{tabular}

    \caption{For gesture recognition using a random forest, computed savings over a traditional array for channel selection, resolution readjustment, and the ratio between resolution readjustment and channel selection.}
    \label{tab:gesture_nl}
\end{table}

\begin{table}[h]
    \centering
    \begin{tabular}{lrrrr}
\toprule
& CS & RR & Ratio & Savings at Eq. Perf. \\
\midrule
0 & 1.778 & 31.508 & 17.721 & 26.913 \\
1 & 5.053 & 11.130 & 2.203 & 1.275 \\
2 & 3.840 & 44.765 & 11.658 & 11.658 \\
3 & 4.800 & 26.597 & 5.541 & 13.853 \\
4 & 2.341 & 35.310 & 15.083 & 15.080 \\
\bottomrule
\end{tabular}

    \caption{For force regression using a neural network, computed savings over a traditional array for channel selection, resolution readjustment, and the ratio between resolution readjustment and channel selection.}
    \label{tab:grasp_nl}
\end{table}

\clearpage

\printbibliography